\documentclass[aps,prb,twocolumn,superscriptaddress,
groupedaddress,10pt]{revtex4}
\usepackage{amsmath}
\usepackage{amssymb}
\usepackage{graphicx}
\usepackage{epsfig}
\usepackage{dcolumn}
\usepackage{bm}
\usepackage{bm}
\usepackage{blindtext}
\usepackage{natbib}
\usepackage{siunitx} 
\usepackage{multirow}

\usepackage[titletoc]{appendix}

\usepackage{simplewick}

\usepackage{bbm}
\makeatletter
\newcommand{\mathleft}{\@fleqntrue\@mathmargin0pt}
\newcommand{\mathcenter}{\@fleqnfalse}
\makeatother

\usepackage[usenames,dvipsnames]{xcolor}
\usepackage{amsthm}
\usepackage{booktabs}
\usepackage{hyperref}
\usepackage{tikz}
\usetikzlibrary{calc}
\usepackage[printwatermark]{xwatermark}
\usepackage{mathtools}

\def\be{\begin{equation}} \def\ee{\end{equation}}
\def\bea{\begin{eqnarray}} \def\eea{\end{eqnarray}}

\def\nn{\nonumber}

\begin{document}
\title{
Symmetry analysis of  bond-alternating Kitaev spin chains and ladders
}

\author{Wang Yang}
\affiliation{Department of Physics and Astronomy and Stewart Blusson Quantum Matter Institute,
University of British Columbia, Vancouver, B.C., Canada, V6T 1Z1}

\author{Alberto Nocera}
\affiliation{Department of Physics and Astronomy and Stewart Blusson Quantum Matter Institute, 
University of British Columbia, Vancouver, B.C., Canada, V6T 1Z1}

\author{Paul Herringer}
\affiliation{Department of Physics and Astronomy and Stewart Blusson Quantum Matter Institute, 
University of British Columbia, Vancouver, B.C., Canada, V6T 1Z1}

\author{Robert Raussendorf}
\affiliation{Department of Physics and Astronomy and Stewart Blusson Quantum Matter Institute, 
University of British Columbia, Vancouver, B.C., Canada, V6T 1Z1}

\author{Ian Affleck}
\affiliation{Department of Physics and Astronomy and Stewart Blusson Quantum Matter Institute, 
University of British Columbia, Vancouver, B.C., Canada, V6T 1Z1}

\begin{abstract}

In this work, we analyze the nonsymmorphic symmetry group structures for a variety of generalized Kitaev spin chains and ladders with bond alternations, including Kitaev-Gamma chain, Kitaev-Heisenberg-Gamma chain, beyond nearest neighbor interactions, and two-leg spin ladders. 
The symmetry analysis is applied to determine the symmetry breaking patterns of several magnetically ordered phases in the bond-alternating Kitaev-Gamma spin chains, as well as the dimerization order parameters for spontaneous dimerizations.
Our work is useful in understanding  the magnetic phases in related models  and may provide  guidance for the symmetry classifications of mean field solutions in further investigations.

\end{abstract}
\maketitle

\section{Introduction}

The honeycomb Kitaev spin-1/2 model \cite{Kitaev2006} is an exactly solvable model which hosts spin liquid ground state as well as fractional excitations including Majorana fermions and nonabelian anyons\cite{Nayak2008}. 
Since the seminal work of Jackeli and Khaliullin\cite{Jackeli2009} in which a realistic method for realizing the Kitaev model in solid state systems is proposed, the field of Kitaev materials  has grown into an intensely studied research area in the past decade \cite{Chaloupka2010,Singh2010,Price2012,Singh2012,Plumb2014,Kim2015,Winter2016,Baek2017,Leahy2017,Sears2017,Wolter2017,Zheng2017,Rousochatzakis2017,Kasahara2018},
and fruitful results  have been obtained. 
However, real materials typically involve additional non-Kitaev interactions\cite{Chaloupka2010,Rau2014,Ran2017,Wang2017,Catuneanu2018,Gohlke2018}, which spoil the exact solvability of the model
and in many cases favor magnetically ordered rather than spin liquid ground states \cite{Liu2011,Chaloupka2013,Plumb2014,Johnson2015}. 
Symmetries typically play important roles in the theoretical studies of the generalized Kitaev models,
including the determinations of symmetry breaking patterns for the ordered phases, the derivations of low energy field theories, and the classifications of mean field solutions in the approach of projective symmetry groups\cite{Wen2002,You2012}. 
Besides the spin-1/2 case, higher spin Kitaev models have also been actively investigated, due to their relevance with real magnetic materials \cite{Baskaran2008,Oitmaa2018,Rousochatzakis2018,Koga2018,Stavropoulos2019,Xu2020}.

From a theoretical point of view, there are intrinsic difficulties in  studying generic strongly correlated two-dimensional (2D) systems both analytically and numerically. 
Therefore, investigations in reduced dimensionality,  i.e., in one-dimensional (1D) systems, can be useful and valuable to better understand the 2D physics, since there are more controllable theoretical tools in 1D \cite{Haldane1981,Haldane1981a,Witten1984,Belavin1984,Knizhnik1984,Affleck1990,White1992,White1993,Schollwock2011}.
Recently, there has been a series of theoretical works on the phase diagrams of 1D generalized Kitaev models \cite{Agrapidis2018,Agrapidis2019,Catuneanu2019,Yang2020,Yang2021,Yang2020b,Yang2021b,Luo2021,Luo2021b,You2020,Sorensen2021}. 
A plethora of interesting phases has been found, including emergent conformal invariance, Luttinger liquid phases, magnetically ordered phases, nonlocal string orders, and spin liquids. 

On the other hand, bond-alternation is a ubiquitous phenomenon in 1D spin systems, which can be induced by the spin-Peierls transition via the coupling with phonons. 
The bond-alternating Kitaev-Gamma chain has been studied for the spin-1/2 and spin-1 cases \cite{Luo2021,Luo2021b},
both exhibiting a rich phase diagram.
However, the structures of the symmetry groups for bond-alternating  Kitaev spin models have not been analyzed before. 

In this work, we perform a systematic study on the symmetry groups of generalized Kitaev spin chains and ladders with bond alternations, from Kitaev-Gamma to Kitaev-Heisenberg-Gamma models, from nearest neighbor to beyond nearest neighbor interactions, and from chains to two-leg ladders.
We find rich  nonsymmorphic symmetry group structures which involve compositions of spatial operations and global spin rotations. 
In particular, with the help of the six-sublattice rotation\cite{Yang2020}, the symmetry group $G$ of a bond-alternating Kitaev-Gamma chain is found to satisfy $G/\mathopen{<} T_{6a}\mathclose{>} \cong O_h$, where $O_h$ is the full octahedral group and $\mathopen{<} T_{6a}\mathclose{>}$ represents the group generated by translation of six lattice sites. 
We note that because of the nonsymmorphic structure, $G$ is not a semi-direct product of $O_h$ and $\mathopen{<} T_{6a}\mathclose{>}$.
Instead, $G$ is a nontrivial group extension of $O_h$ by $\mathopen{<} T_{6a}\mathclose{>}$.

As applications, the symmetries are used to analyze several scenarios.
The symmetry breaking patterns are derived  for the magnetically ordered FM$_{U_6}$, $M_1$ and $M_2$ phases discovered in Ref. \onlinecite{Luo2021}, as well as the FM$_{U_6}$, $M_O$ and $M_I$ phases discovered in Ref. \onlinecite{Luo2021b}:
$O_h\rightarrow D_4$ for FM$_{U_6}$; $O_h\rightarrow D_3$ for $M_O$, $M_1$, $M_2$;
and $O_h\rightarrow D_2$ for $M_I$.
The center of mass spin directions in a unit cell 
for the $O_h\rightarrow D_4$, $O_h\rightarrow D_3$, and $O_h\rightarrow D_2$ symmetry breaking patterns
are found to point towards the face centers, vertices, and edge middles of a spin cube, respectively, 
 in the corresponding six, eight, and twelve-fold degenerate ground states, 
where  $D_n$ is the dihedral group of order $2n$.
In addition, the relations among dimerization order parameters are determined for the spontaneous dimerization, 
and the structure of the symmetry group is analyzed when a nonzero magnetic field is applied along the $(1,1,1)$-direction. 
Our work is useful to better understand various magnetic phases in bond-alternating 1D Kitaev spin models, and may provide  guidance for the symmetry classifications of mean field solutions in further investigations. 

The rest of the paper is organized as follows. 
In Sec. \ref{sec:KG}, a detailed analysis is given to the symmetry aspects of the dimerized Kitaev-Gamma chain,
including the model Hamiltonian, the symmetry group structure, the symmetry breaking patterns of several magnetic phases, and the effect of an external magnetic field.
Sec. \ref{sec:KHG} is devoted to the more general Kitaev-Heisenberg-Gamma model, as well as beyond next nearest neighbor interactions. 
In Sec. \ref{sec:ladder}, the Kitaev spin ladders are analyzed.
Finally, in Sec. \ref{sec:conclusion}, we briefly summarize the results of the paper. 

\section{Bond-alternating Kitaev-Gamma spin chain}
\label{sec:KG}

In this section, we first show that the symmetry group $G$ of a bond-alternating Kitaev-Gamma spin chain satisfies $G/\mathopen{<} T_{6a}\mathclose{>}\cong O_h$,
where $T_{na}$ is the translation by $n$ sites
and $\mathopen{<} ...\mathclose{>}$ represents the group generated by the operations within the brackets.
We also point out that $G$ is a nontrivial group extension of $O_h$ by $\mathopen{<} T_{6a}\mathclose{>}$.
Compared with the uniform Kitaev-Gamma spin chain (i.e., without bond-alternation) whose symmetry group structure has been demonstrated in Ref. \onlinecite{Yang2020} to be $G_u/\mathopen{<} T_{3a}\mathclose{>}\cong O_h$, 
we see that the symmetries are ``halved" in the bond-alternating case. 

Then we discuss a variety of the applications of the symmetry analysis, including derivations of the symmetry breaking patterns of several magnetically ordered phases discovered in Ref. \onlinecite{Luo2021,Luo2021b}, symmetry constraints for dimerization order parameters in spontaneous dimerizations, and the symmetry group in a nonzero magnetic field along $(1,1,1)$-direction.

\subsection{The Hamiltonian}
\label{sec:Hamiltonian}

\begin{figure}[h]
\begin{center}
\includegraphics[width=8.5cm]{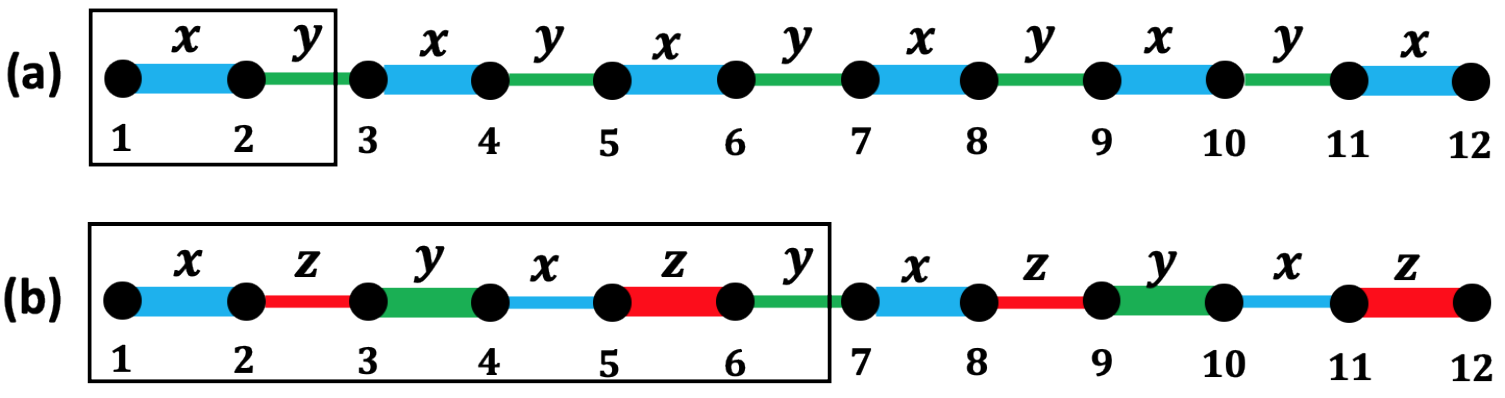}
\caption{Bond patterns of the bond-alternating Kitaev-Gamma chain (a) before the six-sublattice rotation,
(b) after the six-sublattice rotation.
The thick colored lines represent bonds with anisotropy $g_x$, and the thin lines represent $g_y$.
} \label{fig:bonds}
\end{center}
\end{figure}

In accordance with Ref. \onlinecite{Luo2021,Luo2021b}, the Hamiltonian of a bond-alternating Kitaev-Gamma spin chain is defined as
\begin{eqnarray}
H=\sum_{l=1}^{N/2} \big(g_x H^{(x)}_{2l-1,2l}+g_y H^{(y)}_{2l,2l+1}\big),
\label{eq:Ham}
\end{eqnarray}
in which $N$ is system length, $N+1$ is identified with $1$, and
\begin{eqnarray}
H^{(\gamma)}_{i,j}=KS_i^{\gamma}S_j^{\gamma}+\Gamma (S_i^{\alpha}S_j^{\beta}+S_i^{\beta}S_j^{\alpha}),
\label{eq:H_gamma}
\end{eqnarray}
where $\alpha\neq \beta\in \{x,y,z\}$ are the two directions other than $\gamma$.
The pattern for the bond $\gamma$ is shown in Fig. \ref{fig:bonds} (a), which alternates between $x$ and $y$.
When $g_x=g_y$, Eq. (\ref{eq:Ham}) reduces to the isotropic Kitaev-Gamma model studied in Ref. \onlinecite{Yang2020}.
We note that there are two free parameters $\theta$ and $g$ in $H$, defined by
\begin{eqnarray}
K=\cos(\theta),~\Gamma=\sin(\theta),~ g=g_y/g_x.
\end{eqnarray}

A useful transformation is the six-sublattice rotation $U_6$, 
which is defined as \cite{Yang2020}
\begin{eqnarray}
\text{Sublattice $1$}: & (x,y,z) & \rightarrow (x^{\prime},y^{\prime},z^{\prime}),\nn\\ 
\text{Sublattice $2$}: & (x,y,z) & \rightarrow (-x^{\prime},-z^{\prime},-y^{\prime}),\nn\\
\text{Sublattice $3$}: & (x,y,z) & \rightarrow (y^{\prime},z^{\prime},x^{\prime}),\nn\\
\text{Sublattice $4$}: & (x,y,z) & \rightarrow (-y^{\prime},-x^{\prime},-z^{\prime}),\nn\\
\text{Sublattice $5$}: & (x,y,z) & \rightarrow (z^{\prime},x^{\prime},y^{\prime}),\nn\\
\text{Sublattice $6$}: & (x,y,z) & \rightarrow (-z^{\prime},-y^{\prime},-x^{\prime}),
\label{eq:6rotation}
\end{eqnarray}
in which "Sublattice $i$" ($1\leq i \leq 6$) represents all the sites $\{i+6n| n\in \mathbb{Z}\}$ in the chain, and $S^\alpha$ ($S^{\prime \alpha}$) is abbreviated as $\alpha$ ($\alpha^\prime$) for short ($\alpha=x,y,z$).
Unlike the uniform case where the transformed Hamiltonian has a three-site periodicity,
 the Hamiltonian $H$ in Eq. (\ref{eq:Ham}) is six-site-periodic after the $U_6$ transformation. 
Denoting $H^\prime$ to be the transformed Hamiltonian,
we have
\begin{eqnarray}
H^\prime=&&\sum_{\langle ij\rangle\in\gamma\,\text{bond}} (g_0+(-)^{i-1}\delta)\nn\\
&& \cdot \big[ -KS_i^\gamma S_j^\gamma -\Gamma (S_i^\alpha S_j^\alpha+S_i^\beta S_j^\beta)\big],
\label{eq:Ham6}
\end{eqnarray}
in which $\langle ij\rangle$ denotes the nearest neighboring bond connecting sites $i$ and $j$, and
\bea
g_0=\frac{g_x+g_y}{2},~\delta=\frac{g_x-g_y}{2}.
\eea
The bond structure for $H^\prime$ is shown in Fig. \ref{fig:bonds} (b),
which shows a screw pattern going as $x$-$z$-$y$.
Appendix \ref{app:6rotation} explicitly displays the terms of $H^\prime$ in a six-site unit cell.

Here we make some comments on the density matrix renormalization group (DMRG) numerics that we employ. 
We solve numerically the dimerized Kitaev-Gamma chain using the DMRG method\cite{White1992,White1993} on a chain of a length up to 144 sites for the spin-1/2 and up to 48 sites for the spin-1 cases,
both with periodic boundary conditions. 
Several sweeps are performed  to reach convergence using up to $m=800$ DMRG states keeping a truncation error below $10^{-6}$.

\subsection{Symmetries}
\label{sec:symmetry}

\begin{figure}[h]
\begin{center}
\includegraphics[width=6cm]{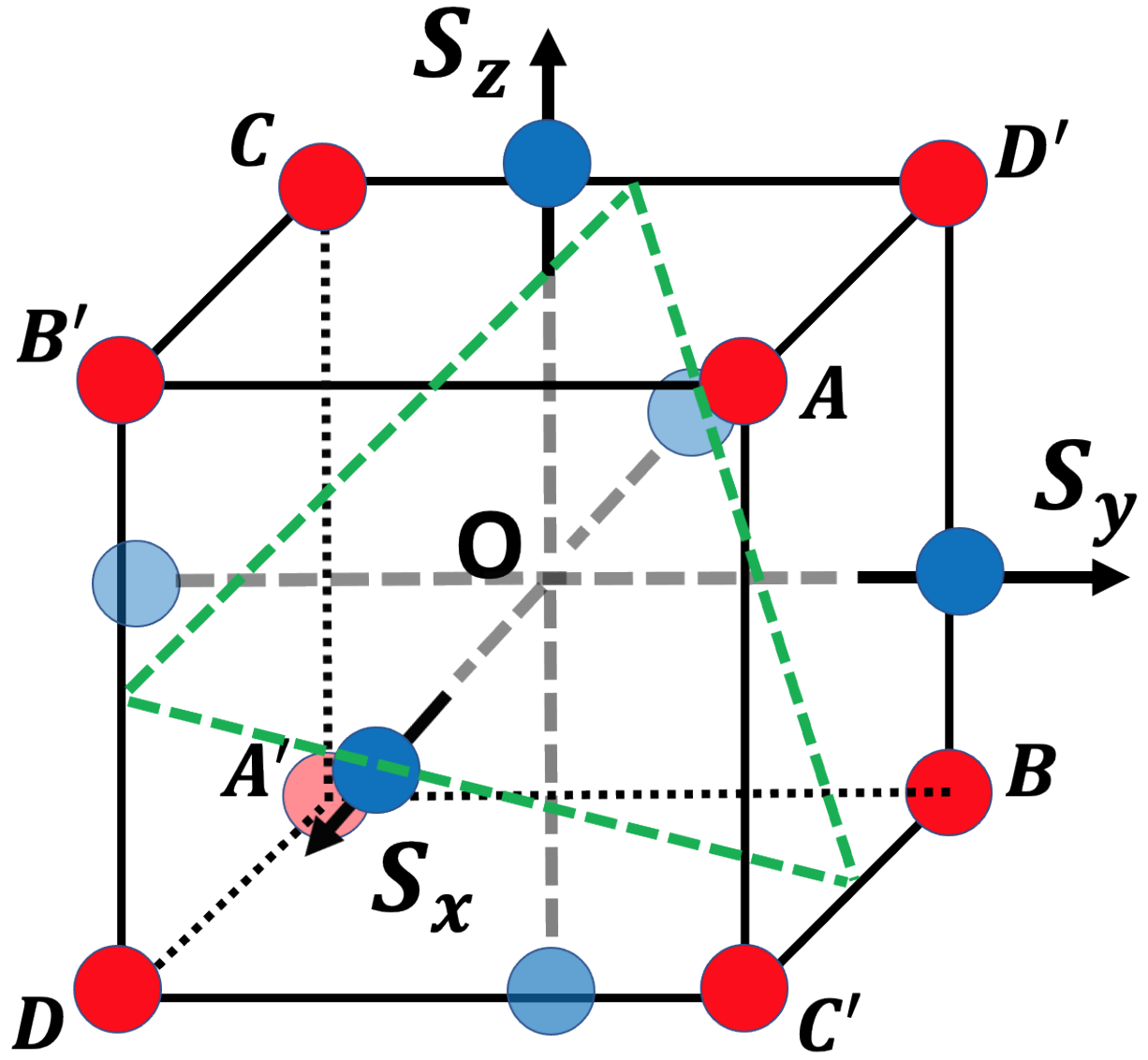}
\caption{Actions of the symmetry transformations in  spin space as symmetry operations of a spin cube.
} \label{fig:Oh}
\end{center}
\end{figure}

It can be verified that the transformed Hamiltonian in Eq. (\ref{eq:Ham6}) is invariant under the following transformation:
\begin{eqnarray}
1.&T &:  (S_i^x,S_i^y,S_i^z)\rightarrow (-S_{i}^x,-S_{i}^y,-S_{i}^z)\nn\\
2. & R_a^{-1} T_{2a} &: (S_i^x,S_i^y,S_i^z)\rightarrow (S_{i+2}^y,S_{i+2}^z,S_{i+2}^x)\nn\\
3.& R_I I T_a&:  (S_i^x,S_i^y,S_i^z)\rightarrow (-S_{7-i}^z,-S_{7-i}^y,-S_{7-i}^x)\nn\\
4. & R(\hat{x},\pi) &: (S_i^x,S_i^y,S_i^z)\rightarrow (S_i^x,-S_i^y,-S_i^z)\nn\\
5. & R(\hat{y},\pi) &: (S_i^x,S_i^y,S_i^z)\rightarrow (-S_i^x,S_i^y,-S_i^z)\nn\\
6. & R(\hat{z},\pi) &: (S_i^x,S_i^y,S_i^z)\rightarrow (-S_i^x,-S_i^y,S_i^z),
\label{eq:symmetry}
\end{eqnarray}
in which: 
$T$ is the time reversal operation; 
$R(\hat{\alpha},\pi)$ ($\alpha=x,y,z$) are  global spin rotations around $\hat{\alpha}$-directions which form a $\mathbb{Z}_2\times\mathbb{Z}_2$ subgroup of the whole symmetry group; 
$T_{na}$ ($n\in \mathbb{Z}$) is the translation operator by $n$ lattice sites where ``$a$" represents the lattice constant; 
$I$ is the inversion operator with respect to site 4, or alternatively, we can write $I T_a$ as $I^\prime$ where $I^\prime$ is the inversion operation with respect to the center of the bond between sites 3 and 4; 
$R_I=R(\hat{n}_I,\pi)$ is a $\pi$-rotation around the $\hat{n}_I=(1,0,-1)/\sqrt{2}$ direction; 
and $R_a=R(\hat{n}_a,2\pi/3)$ is a $2\pi/3$-rotation around the $\hat{n}_a=(1,1,1)/\sqrt{3}$ direction. 
In general, $R(\hat{n},\phi)$ is used to denote a global spin rotation around $\hat{n}$-direction by an angle $\phi$.
The symmetry group $G$ is generated by the above transformations, i.e.,
\begin{eqnarray}
G=\mathopen{<}T, R_I I T_a,R_a^{-1} T_{2a},R(\hat{x},\pi),R(\hat{y},\pi),R(\hat{z},\pi)\mathclose{>}.
\end{eqnarray}

As can be seen from Fig. \ref{fig:Oh}, if we ignore the spatial components involving $I, T_a$ in Eq. (\ref{eq:symmetry}),
then the actions of all symmetry transformations restricted within the spin space are symmetry transformations of  the three-dimensional spin cube.
In particular, the time reversal operation can be viewed as inversion within the spin space,
since it changes the sign of the spin operators.
The above observation indicates that the symmetry group $G$ is closely related to the $O_h$ group, which is the symmetry group of a cube.
Indeed, we are able to show that  
\begin{eqnarray}
G/\mathopen{<} T_{6a}\mathclose{>}\cong O_h,
\end{eqnarray}
which will be proved in Sec. \ref{sec:proof}.

Notice that six is the smallest period for the spatial translation symmetry,
hence $G/\mathopen{<}T_{na}\mathclose{>}$ is meaningful only for $n\in 6\mathbb{Z}$.
In contrast, the period is three for a uniform Kitaev-Gamma chain (i.e., $g_x=g_y$),  
and as a result, $G_u/\mathopen{<} T_{3a}\mathclose{>}\cong O_h$ in the uniform case, where $G_u$ is the symmetry group for $g_x=g_y$.
Comparing the group structures of $G$ with $G_u$,
we see that the symmetries in $G$ are halved with respect to $G_u$.
This is intuitively correct since an alternation of stronger and weaker bonds  is introduced. 

\subsection{Proof of $G/\mathopen{<} T_{6a}\mathclose{>} \cong O_h$}
\label{sec:proof}

We give a proof for the group structure $G/\mathopen{<} T_{6a}\mathclose{>}\cong O_h$.
The method is similar with the analysis in Ref. \onlinecite{Yang2020}, where $G_u/\mathopen{<} T_{3a}\mathclose{>}\cong O_h$ is proved for the uniform  Kitaev-Gamma chain.

The proof consists of two parts, including $G/\mathopen{<} T_{6a}\mathclose{>} \subseteq O_h$ and $|O_h| \leq |G/\mathopen{<} T_{6a}\mathclose{>}|$, where  $|A|$ represents the order (i.e., number of elements) of the group $A$.

\subsubsection{$G/\mathopen{<} T_{6a}\mathclose{>} \subseteq O_h$}

There is a generator-relation representation for the $O_h$ group\cite{Coxeter1965}:
\begin{flalign}
O_h=\mathopen{<} r,s,t| r^2=s^2=t^2=(rs)^3=(st)^4=(rt)^2=e \mathclose{>},
\label{eq:Generator_Oh}
\end{flalign}
in which $e$ is the identity element, and the geometrical meanings of the generators $r,s,t$  are three reflections.
We are going to construct $r,s,t$ out of $\{T,R_a^{-1}T_{2a},R_I IT_a, R(\hat{x},\pi),R(\hat{y},\pi),R(\hat{z},\pi)\}$,
and verify the following two Assertions: 
\begin{enumerate}
\item \label{assersion1} $r,s,t$ satisfy the relations in Eq. (\ref{eq:Generator_Oh}) in the sense of modulo $T_{6a}=(rs)^3$;
\item\label{assertion2} all symmetry operations in Eq. (\ref{eq:symmetry}) can be obtained from $r,s,t$.
\end{enumerate}
Notice that Assertion \ref{assersion1}  ensures that  $\mathopen{<}r,s,t\mathclose{>}/\mathopen{<} T_{6a}\mathclose{>}$  is a subgroup of $O_h$,
whereas Assertion \ref{assertion2} ensures that $G$ is equal to $\mathopen{<}r,s,t\mathclose{>}$.
Then $G/\mathopen{<} T_{6a}\mathclose{>}$ has to be a subgroup of $O_h$ as a result of the above two assertions. 

The generators can be constructed as follows, 
\begin{flalign}
&r =T\cdot R_I IT_{a}\cdot T_{-6a}, \nn\\
&s =T\cdot (R_a^{-1}T_{2a})\cdot R_I IT_a\cdot (R_a^{-1}T_{2a})^{-1},  \nn\\
&t= T \cdot R(\hat{y},\pi). 
\label{eq:generators}
\end{flalign}
In spin space, their actions are given by
\begin{flalign}
\begin{array}{ c c  c }
r& (x,y,z)\rightarrow (z,y,x) & \text{Reflection to $ABA^{\prime}B^{\prime}$-plane} \\
s& (x,y,z)\rightarrow (y,x,z) & \text{Reflection to $ACA^{\prime}C^{\prime}$-plane} \\
t& (x,y,z)\rightarrow (x,-y,z) & \text{Reflection to $xz$-plane}, 
\end{array}
\end{flalign}
in which the second and the third columns show the actions in the spin space (where $S^\alpha$ is denoted as $\alpha$ for short)
and the geometrical meanings as symmetries of a cube in Fig. \ref{fig:Oh}, respectively.
It can be proved that
\begin{eqnarray}
r^2=s^2=t^2=(st)^4=(rt)^2=1,~(rs)^3=T_{6a}.
\label{eq:relations_i}
\end{eqnarray}
Detailed verifications  of the relations in Eq. (\ref{eq:relations_i}) are included in Appendix \ref{app:verify}.
Notice that $T_{6a}=(rs)^3$  belongs to the group generated by $\{r,s,t\}$,
and it can be easily seen that $\mathopen{<} T_{6a}\mathclose{>}$ is a normal subgroup.
Hence  the quotient $\mathopen{<}r,s,t\mathclose{>}/\mathopen{<} T_{6a}\mathclose{>}$ forms a group.
Then Eq. (\ref{eq:relations_i}) implies Assertion 1, and as a consequence, $\mathopen{<}r,s,t\mathclose{>}/\mathopen{<} T_{6a}\mathclose{>}\subseteq O_h$.

Next we verify Assertion \ref{assertion2}. 
The constructions go as follows:
\begin{eqnarray}
T&=&(rs)^{-1}(st)^2 r (st)\nn\\
R_IIT_a&=&(st)^2rsts(rs)^3\nn\\
R_a^{-1}T_{2a}&=&rs\nn\\
R(\hat{x},\pi)&=&r(ts)^2r\nn\\
R(\hat{y},\pi)&=&sr(st)^2rs\nn\\
R(\hat{z},\pi)&=&(st)^2,
\label{eq:symmetry_from_rst}
\end{eqnarray}
which demonstrates that $G=\mathopen{<}r,s,t\mathclose{>}$.
The verifications of Eq. (\ref{eq:symmetry_from_rst}) are included in Appendix \ref{app:verify_B}.

\subsubsection{$|O_h| \leq |G/\mathopen{<} T_{6a}\mathclose{>}|$}

\begin{table}
\begin{center}
\begin{tabular}{| c | c | c | c | c | c |}
\hline
 $E$ & $1$ & $(x,y,z)$ & $\mathbbm{1}$ & $\mathbbm{1}$ & e  \\
 \hline
 \multirow{3}{*}{$3C_2$}& $2$ & $(x,-y,-z)$ & $R(OX,\pi)$ & $\mathbbm{1}$ & $r(ts)^2r$\\
 \cline{2-6}
& $3$ & $(-x,y,-z)$ & $R(OY,\pi)$ & $\mathbbm{1}$ & $sr(st)^2rs$ \\
 \cline{2-6}
& $4$ & $(-x,-y,z)$ & $R(OZ,\pi)$ & $\mathbbm{1}$ & $(st)^2$ \\
 \hline
 \multirow{6}{*}{$6C_4$}  & $5$ & $(x,z,-y)$ & $R(OX,\frac{\pi}{2})$ & $IT_{-7a}$ & $trsr$ \\
 \cline{2-6}
 & $6$ & $(x,-z,y)$ & $R(OX^{\prime},\frac{\pi}{2})$ & $IT_{-7a}$ & $rsrt$ \\
 \cline{2-6}
 & $7$ & $(-z,y,x)$ & $R(OY,\frac{\pi}{2})$ & $IT_{-5a}$ & $rsts$ \\
 \cline{2-6}
 & $8$ & $(z,y,-x)$ & $R(OY^{\prime},\frac{\pi}{2})$ & $IT_{-5a}$ & $stsr$ \\
 \cline{2-6}
 & $9$ & $(y,-x,z)$ & $R(OZ,\frac{\pi}{2})$ & $IT_{-3a}$ & $st$\\
 \cline{2-6}
 & $10$ & $(-y,x,z)$ & $R(OZ^{\prime},\frac{\pi}{2})$ & $IT_{-3a}$ & $ts$ \\
 \hline
 \multirow{6}{*}{$6C_{2}^{'}$} & $11$ & $(y,x,-z)$ & $R([AC],\pi)$ & $IT_{-5a}$ & $rsrtsr$ \\
 \cline{2-6}
 & $12$ & $(-y,-x,-z)$ & $R([BD],\pi)$ & $IT_{-3a}$ & $r(ts)^2rst$\\
 \cline{2-6}
 & $13$ &$(z,-y,x)$ & $R([AB],\pi)$  & $IT_{-5a}$ & $rt$\\
 \cline{2-6}
 & $14$ & $(-z,-y,-x)$ & $R([CD],\pi)$ & $IT_{-5a}$ & $(st)^2rsts$\\
 \cline{2-6}
 & $15$ & $(-x,z,y)$ & $R([AD],\pi)$ & $IT_{-a}$ & $strs$\\
 \cline{2-6}
 & $16$ & $(-x,-z,-y)$ & $R([BC],\pi)$ & $IT_{-a}$ & $tsrtst$\\
 \hline
 \multirow{8}{*}{$8C_3$}&  $17$ & $(y,z,x)$ &  $R(OA,\frac{2\pi}{3})$ & $T_{2a}$ & $rs$ \\
 \cline{2-6}
 & $18$ & $(z,x,y)$ & $R(OA^{\prime},\frac{2\pi}{3})$ & $T_{-2a}$ & $sr$\\
 \cline{2-6}
 & $19$ & $(-y,-z,x)$ & $R(OB,\frac{2\pi}{3})$ & $T_{2a}$ & $trst$ \\
 \cline{2-6}
 & $20$ & $(z,-x,-y)$ &  $R(OB^{\prime},\frac{2\pi}{3})$ & $T_{-2a}$ & $tsrt$ \\
 \cline{2-6}
 & $21$ & $(y,-z,-x)$ &  $R(OC,\frac{2\pi}{3})$ & $T_{2a}$ & $stsrst$\\
 \cline{2-6}
 & $22$ & $(-z,x,-y)$ &  $R(OC^{\prime},\frac{2\pi}{3})$ & $T_{-2a}$ & $tsrsts$\\
 \cline{2-6}
 & $23$ & $(-y,z,-x)$  &  $R(OD,\frac{2\pi}{3})$ & $T_{2a}$ & $(st)^2rs$ \\
 \cline{2-6}
 & $24$ & $(-z,-x,y)$ &  $R(OD^{\prime},\frac{2\pi}{3})$ & $T_{-2a}$ & $sr(ts)^2$ \\
 \hline
\end{tabular}
\caption{List of $24$ the proper group elements of the point group $O_h$, which form the group $O$.
The 24 improper elements in the group $O_h$ can be obtained from the elements in this table by multiplying with the time reversal operation, i.e., $(rs)^{-1}(st)^2 r (st)$.
In accordance with the notations in Fig. \ref{fig:Oh}, $OM$ represents the vector pointing from the center of the cube (i.e. the point $O$) to the vertex or the direction $M$, where $M$ is one of $A,\,A^{\prime},\,B,\,B^{\prime},\,C,\,C^{\prime},\,D,\,D^{\prime}$ when it is a vertex of the cube, or is one of $X,\,Y,\,Z,\,X^{\prime},\,Y^{\prime},\,Z^{\prime}$ when it represents a direction.
$X,\,Y,\,Z$ represent the positive directions of the three axes $x,\,y,\,z$, and $X^{\prime},\,Y^{\prime}\,Z^{\prime}$ represent the negative directions of the three axes.
The symbol $[MN]$ represents the line passing through the point that bisects the edge $MN^{\prime}$ and the point that bisects $M^{\prime}N$,
where $M,\,N,\,M^{\prime},\,N^{\prime}$ are all vertices of the cube.
The third, fourth, fifth and sixth  columns give the action in the spin space, the geometrical meaning, the spatial component, and the expression in terms of the generators, for each symmetry operation, respectively. 
\label{table:O}
}
\end{center}
\end{table}

To further prove $O_h\cong G/\mathopen{<} T_{6a}\mathclose{>}$, it is enough to show that $|O_h|\leq|G/\mathopen{<} T_{6a}\mathclose{>}|$
since we have already shown that the latter is a subgroup of the former.
For this, we will construct 48 distinct elements within $G/\mathopen{<} T_{6a}\mathclose{>}$.
Using the fact that $|O_h|=48$, the statement follows directly. 

The construction of 24 distinct elements in $G/\mathopen{<} T_{6a}\mathclose{>}$ is shown in the last column in Table \ref{table:O}.
The third and the fourth columns both give the actions of the corresponding transformations restricted within the spin space,
and the fifth column give the spatial components of the symmetries.
The expressions in the table can be checked using the similar method as the one used in Appendix \ref{app:verify_B}.

Since the 24 elements all have distinct actions in the spin space, they of course must be different when spatial components are also taken into account.
The other 24 elements can be obtained by multiplying those in Table \ref{table:O} with time reversal operation $T=(rs)^{-1}(st)^2 r (st)$.
Since the determinants of the corresponding matrices of the linear transformations in spin space become $-1$ after composing with $T$, the newly obtained 24 elements must be different from the older ones.
This proves that $|G/\mathopen{<} T_{6a}\mathclose{>}|\geq 48$.

\subsection{Group extension} 
\label{sebsec:group_extension}

In this subsection, we make some further comments on the group structure of the symmetry group $G$.
Although $G/\mathopen{<} T_{6a}\mathclose{>}\cong O_h$, we will show that $G$ is not a semi-direct product of $O_h$ and $\mathopen{<} T_{6a}\mathclose{>}$.
In fact, $G$ corresponds to a nontrivial group extension of $O_h$ by $\mathopen{<} T_{6a}\mathclose{>}$.

First, we briefly review some basic facts about group extensions \cite{Hiller1986}.
Consider  the following short exact sequence,
\bea
1\xrightarrow[]{} N \xrightarrow[]{i} G_0 \xrightarrow[]{\pi} H \rightarrow 1
\label{eq:short_exact}
\eea
in which $1$ represents the trivial group containing a single element, the arrows represent group homomorphisms,  $N$ is an abelian group,  $i$ and $\pi$ are group homomorphisms which are injective and surjective, respectively, 
and $\text{Ker}(\pi)=\text{Im}(i)$.
Given Eq. (\ref{eq:short_exact}), $G_0$ is called a group extension of $H$ by $N$.
Since the homomorphism $i$ is injective, $N$ can be naturally viewed as a subgroup of $G_0$ via the embedding induced by $i$.
Furthermore, since  $\text{Ker}(\pi)=\text{Im}(i)\cong N$ and $\pi$ is surjective, $N$ is a normal subgroup of $G_0$ and $H\cong G_0/N$.
As a result, there is a natural group action $\varphi$ of $H$ on $N$, defined as
\bea
\varphi_h(n)=x_hn(x_h)^{-1},
\eea
in which $h\in H$, $n\in N$, and $x_h\in G_0$ satisfies $\pi(x_h)=h$.
In this way, $N$ becomes an $H$-module via $\varphi$.
Notice that although the choice of $x_h$ is not unique (up to some element in $N$),
$\varphi_h$ is well-defined  since $N$ is abelian.

Eq. (\ref{eq:short_exact}) determines how $G_0$ is formed out of $H$ and $N$.
A simple example is $G_0$ as a semi-direct product of $H$ and $N$, i.e., $G_0 \cong N\rtimes_{\varphi} H$, where $\varphi$ is the group action under consideration. 
In this case, $H$ can be naturally viewed as a subgroup of $G_0$,
or equivalently, there exists an injective group homomorphism $\tau:H\rightarrow G_0$ such that $\pi\cdot \tau=id$ where ``$\cdot $" represents the composition of maps and $id$ is the identity map on $H$. 
However, in more general situations, $\tau$ may not exist and $G_0$ is not of a semi-direct product structure.
In fact, it is known that group extensions can be classified by the second cohomology group $H^2(H,N)$\cite{Dummit1991},
and semi-direct products correspond to the trivial element in  $H^2(H,N)$.
A brief review for the relation between group extensions and $H^2(H,N)$ is given in Appendix \ref{app:cohomology}.

Coming back to the case of dimerized Kitaev-Gamma model, we have the short exact sequence 
\bea
1\xrightarrow[]{} \mathopen{<} T_{6a}\mathclose{>} \xrightarrow[]{i} G \xrightarrow[]{\pi} O_h \rightarrow 1.
\eea
We will show that there does not exist any group homomorphism $\tau:O_h\rightarrow G$ such that $\pi\cdot \tau=id$,
thereby negating the semi-direct product structure. 
We prove by contradiction.
Suppose otherwise such a group homomorphism $\tau$ exists.
Consider $R_a^{-1}T_{2a}\in G$, and let $y=\pi(R_a^{-1}T_{2a})$.
Since $\pi(\tau(y))=\pi(R_a^{-1}T_{2a})=y$,  $\tau(y)$ and $R_a^{-1}T_{2a}$ must only differ by an element in $\text{Ker}(\pi)=\mathopen{<} T_{6a}\mathclose{>}$, i.e., there exists $n\in \mathbb{Z}$, such that $\tau(y)=R_a^{-1}T_{(2+6n)a}$.
Since $\tau$ is assumed to be a homomorphism, we have
\bea
\tau(y^3)=[\tau(y)]^3=T_{(1+3n)6a}.
\label{eq:y3}
\eea
Notice that $T_{(1+3n)6a}\in \text{Ker}(\pi)$, hence
\bea
\pi(\tau(y^3))=e,
\eea
where $e$ is the identity element in $O_h$.
On the other hand, recall that $\pi\cdot \tau=1$, thus $y^3=e$.
However, for $\tau$ to be a group homomorphism, we must have
\bea
\tau(y^3)=\tau(e)=e_0,
\label{eq:y3_b}
\eea
where $e_0$ is the identity element in $G$.
Clearly, Eq. (\ref{eq:y3_b}) contradicts with Eq. (\ref{eq:y3}), since $1+3n$ can never be $0$ (so that $T_{(1+3n)6a}\neq e_0$ for whatever $n\in \mathbb{Z}$).

The above result can be understood from the point of view of the second cohomology group.
Since $\mathopen{<}R_a^{-1}T_{2a}\mathclose{>}$ is a subgroup of $G$, we may consider the following exact sequence
\bea
1\xrightarrow[]{} \mathopen{<} T_{6a}\mathclose{>} \xrightarrow[]{i} \mathopen{<}R_a^{-1}T_{2a}\mathclose{>} \xrightarrow[]{\pi} C_3 \rightarrow 1,
\label{eq:exact_C3}
\eea
in which $C_3\simeq \mathbb{Z}_3(=\mathbb{Z}/3\mathbb{Z})$ is the cyclic group of order three, a subgroup of $O_h$ obtained from the image of $ \mathopen{<}R_a^{-1}T_{2a}\mathclose{>}$ under the homomorphism $\pi$.
A cleaner way to write Eq. (\ref{eq:exact_C3}) is
\bea
1\xrightarrow[]{} \mathbb{Z} \xrightarrow[]{i} \frac{1}{3}\mathbb{Z} \xrightarrow[]{\pi} \mathbb{Z}_3 \rightarrow 1.
\eea
Notice that in this case, since $R_a^{-1}T_{2a}$ commutes with $T_{6a}$, the group action of $\mathbb{Z}_3$ on $\mathbb{Z}$ is trivial, i.e., $\varphi_a(b)=b$ for any $a\in \mathbb{Z}_3, b\in \mathbb{Z}$.
On the other hand, it is known that $H^2(\mathbb{Z}_3,\mathbb{Z})\cong \mathbb{Z}_3$ for such  trivial group action \cite{Chen2013}.
Indeed, the dimerized Kitaev-Gamma chain corresponds to a nontrivial element in $H^2(\mathbb{Z}_3,\mathbb{Z})$.

\subsection{Symmetry breaking patterns of magnetically ordered phases}
\label{sec:Kee_phases}

We apply the previous symmetry analysis to several magnetically ordered phases in the spin-$S$ bond-alternating Kitaev-Gamma chain.
We will figure out the symmetry breaking patterns for the FM$_{U_6}$, $M_1$ and $M_2$ phases in the spin-1/2 case discovered in Ref. \onlinecite{Luo2021}, as well as the FM$_{U_6}$, $M_O$ and $M_I$ phases in the spin-1 case in Ref. \onlinecite{Luo2021b}.
Three different types of symmetry breaking patterns are found, including $O_h\rightarrow D_4$, $O_h\rightarrow D_3$, and $O_h\rightarrow D_2$.

\subsubsection{The FM$_{U_6}$ phase: $O_h\rightarrow D_4$}

The FM$_{U_6}$ phase appears in both spin-$1/2$ and spin-$1$ cases.
The spin orderings in one of the degenerate ground states in the FM$_{U_6}$ phase are given by
\begin{flalign}
&\langle \vec{S}_{1+6n}\rangle=c\hat{z},~\langle \vec{S}_{2+6n}\rangle=b\hat{z},~\langle \vec{S}_{3+6n}\rangle=b\hat{z},\nn\\
&\langle \vec{S}_{4+6n}\rangle=c\hat{z},~\langle \vec{S}_{5+6n}\rangle=a\hat{z},~\langle \vec{S}_{6+6n}\rangle=a\hat{z},
\label{eq:spin_order_FMU6}
\end{flalign}
which have been shown numerically in Ref. \onlinecite{Luo2021} for the spin-$1/2$ case, and in Ref. \onlinecite{Luo2021b} for the spin-$1$ case.
We will demonstrate that the unbroken symmetry group of the pattern in Eq. (\ref{eq:spin_order_FMU6}) is 
\begin{eqnarray}
H_1=\mathopen{<}TR(\hat{y},\pi),st,T_{6a}\mathclose{>},
\end{eqnarray}
in which $s$ and $t$  as two of the generators of the full symmetry group are given in Eq. (\ref{eq:generators}).
It can be verified that the expression of $st$ can be simplified as
\begin{eqnarray}
st=R(\hat{z},-\frac{\pi}{2})IT_{-3a}.
\end{eqnarray}

Let's figure out the allowed spin ordering patterns which are invariant under $H_1$.
First notice that $T_{6a}$ belongs to $H_1$, hence the spin alignments are six-site periodic. 
We will focus on a single unit cell in which the site indices should be understood as modulo six.
The actions of the two generators of $H_1$ on the spins are
\begin{eqnarray}
TR(\hat{y},\pi):&\left(\begin{array}{c}
S_i^x\\
S_i^y\\
S_i^z
\end{array}\right)\rightarrow 
\left(\begin{array}{c}
S_i^x\\
-S_i^y\\
S_i^z
\end{array}\right),\nn\\
st:&\left(\begin{array}{c}
S_i^x\\
S_i^y\\
S_i^z
\end{array}\right)\rightarrow 
\left(\begin{array}{c}
-S_{5-i}^y\\
S_{5-i}^x\\
S_{5-i}^z
\end{array}\right).
\end{eqnarray}
Therefore, the invariance under $H_1$ requires
\begin{flalign}
S_i^x=-S_{5-i}^y=0,~S_{5-i}^z=S_i^z.
\end{flalign}
Denoting 
\bea
\langle S_1^z\rangle=\langle S_4^z\rangle=c,~\langle S_2^z\rangle=\langle S_3^z\rangle=b,~\langle S_5^z\rangle=\langle S_6^z\rangle=a,
\eea 
we obtain Eq. (\ref{eq:spin_order_FMU6}).

Next, we prove that $H_1/\mathopen{<} T_{6a}\mathclose{>}$ is isomorphic to $D_4$, where $D_n$ is the dihedral group of order $2n$.
First, we show that $H_1/\mathopen{<} T_{6a}\mathclose{>}$ is a subgroup of $D_4$.
For this, the following generator-relation representation of the $D_n$ group is needed \cite{Coxeter1965}:
\bea
D_n=\mathopen{<}a,b|a^n=b^2=(ab)^2=e\mathclose{>}.
\label{eq:Dn}
\eea
Let $st=a_1$, and $TR(\hat{y},\pi)=b_1$.
It can straightforwardly checked that 
\bea
(a_1)^4=1, ~(b_1)^2=1,~(a_1 b_1)^2=1,
\eea
which are all $e$ in the sense of modulo $T_{6a}$.
Hence the two generators of $H_1/\mathopen{<} T_{6a}\mathclose{>}$ satisfy the relations in Eq. (\ref{eq:Dn}),
and as a result, $H_1/\mathopen{<} T_{6a}\mathclose{>}$ must be a subgroup of $D_4$.
Secondly,  to further prove that $H_1/\mathopen{<} T_{6a}\mathclose{>} \cong D_4$, it is enough to show that $H_1$ contains at least $|D_4|=8$ group elements.
Eight distinct elements can be constructed as
\bea
\{1,a_1,(a_1)^2,(a_1)^3,a_1b_1,(a_1)^2b_1,(a_1)^3b_1\}.
\eea
These eight operations are mutually distinct, since when restricted to the spin space,
they correspond to the eight symmetry operations of the square $AB^\prime CD^\prime$ in Fig. \ref{fig:Oh}. 

The above analysis proves that the symmetry breaking pattern for the FM$_{U_6}$ phase is  $O_h\rightarrow D_4$.
Since $|O_h/D_4|=6$, there are six degenerate ground states. 
The center of mass directions of the six spins in a unit cell point to the $\pm \hat{x}$, $\pm \hat{y}$, and  $\pm \hat{z}$ directions in the six degenerate states, which correspond to the face centers of the cube in Fig. \ref{fig:Oh}.

\subsubsection{The $M_1$, $M_2$ and $M_O$ phases: $O_h\rightarrow D_3$}

The $M_1$ and $M_2$ phases are found in the spin-1/2 bond-alternating Kitaev-Gamma chain \cite{Luo2021}, and the $M_O$ phase is found in the spin-1 case \cite{Luo2021b}, which all exhibit a six-site periodicity in the spin orderings.
The spin expectation values in a unit cell in all these three phases are given by ($\eta_\alpha=\pm 1$, where $\alpha=x,y,z$)
\begin{flalign}
&\langle \vec{S}_1\rangle = \left(\begin{array}{c}
\eta_x a\\
\eta_y b\\
\eta_z c
\end{array}\right),
~\langle \vec{S}_2\rangle = \left(\begin{array}{c}
\eta_x a\\
\eta_y c\\
\eta_z b
\end{array}\right),
~\langle \vec{S}_3\rangle = \left(\begin{array}{c}
\eta_x c\\
\eta_y a\\
\eta_z b
\end{array}\right),\nn\\
&\langle \vec{S}_4\rangle = \left(\begin{array}{c}
\eta_x b\\
\eta_y a\\
\eta_z c
\end{array}\right),
~\langle \vec{S}_5\rangle = \left(\begin{array}{c}
\eta_x b\\
\eta_y c\\
\eta_z a
\end{array}\right),
~\langle \vec{S}_6\rangle = \left(\begin{array}{c}
\eta_x c\\
\eta_y b\\
\eta_z a
\end{array}\right),
\label{eq:M12}
\end{flalign}
which  is numerically shown for the $M_1$ and $M_2$ phases in the spin-$1/2$ case in Ref. \onlinecite{Luo2021}, and for the $M_O$ phase in the spin-$1$ case in Ref. \onlinecite{Luo2021b}.
Clearly, the ground states are eight-fold degenerate, and the center of mass spin directions in a unit cell in the corresponding eight degenerate ground states point to the eight vertices of the spin cube in Fig. \ref{fig:Oh}.
We will demonstrate that in the sense of modulo $T_{6a}$, the residual symmetry group of the spin alignments in Eq. (\ref{eq:M12}) is $D_3$.
Hence the symmetry breaking pattern is $O_h\rightarrow D_3$. 

Consider the following group
\bea
H_2=\mathopen{<}R_a^{-1}T_{2a}, s\mathclose{>},
\label{eq:H2_a}
\eea
in which $s$ is given in Eq. (\ref{eq:generators}).
It has been verified in Eq. (\ref{eq:generators}) that 
$s=M_{AC} IT_{3a}$,
in which $M_{AC}$ is a reflection in the spin space, defined as
$M_{AC}: (x,y,z)\rightarrow (y,x,z)$.

According to Eq. (\ref{eq:symmetry}), the invariance under $R_{a}^{-1}T_{2a}$ requires
\bea
x_{i+2}=z_i,~y_{i+2}=x_i,~z_{i+2}=y_i.
\eea
Hence the spin orderings must be
\begin{flalign}
&\langle \vec{S}_1\rangle = \left(\begin{array}{c}
a\\
b\\
c
\end{array}\right),
~\langle \vec{S}_3\rangle = \left(\begin{array}{c}
c\\
a\\
b
\end{array}\right),
~\langle \vec{S}_5\rangle = \left(\begin{array}{c}
b\\
c\\
a
\end{array}\right),\nn\\
&\langle \vec{S}_2\rangle = \left(\begin{array}{c}
a^\prime\\
b^\prime\\
c^\prime
\end{array}\right),
~\langle \vec{S}_4\rangle = \left(\begin{array}{c}
c^\prime\\
a^\prime\\
b^\prime
\end{array}\right),
~\langle \vec{S}_1\rangle = \left(\begin{array}{c}
b^\prime\\
c^\prime\\
a^\prime
\end{array}\right).
\label{eq:M12_b}
\end{flalign}
The action of $M_{AC}IT_{3a}$ is 
\bea
M_{AC}IT_{3a}: (x_i.y_i,z_i)\rightarrow (y_{5-i},x_{5-i},z_{5-i}).
\eea
Thus, the invariance under $M_{AC}IT_{3a}$ requires
\bea
a^\prime=a,~b^\prime=c,~c^\prime=b,
\eea
which reduces Eq. (\ref{eq:M12_b})  to Eq. (\ref{eq:M12}).

The above  analysis demonstrates that the unbroken symmetry group for the ground state corresponding to $\eta_x=\eta_y=\eta_z=1$  is the group $H_2$ defined in Eq. (\ref{eq:H2_a}).
Next we prove that $H_2/\mathopen{<} T_{6a}\mathclose{>}$ is isomorphic to $D_3$.
Let $a_2=R_a^{-1}T_{2a}$, and $b_2=s$. 
Then clearly
\bea
(a_2)^3=T_{6a},~(b_2)^2=1,~(a_2b_2)^2=1.
\eea
According to Eq. (\ref{eq:Dn}), $H_2/\mathopen{<}T_{6a}\mathclose{>}$ satisfies the generator-relation representation for the $D_3$ group, hence it is a subgroup of $D_3$.
Furthermore, the following six elements are all distinct 
\bea
\{1,a_2,(a_2)^2,b_2,a_2b_2,(a_2)^2b_2\},
\eea
since when restricted to the spin space, they correspond to the six symmetry operations of the green dashed triangle in Fig. (\ref{fig:Oh}).
As a result, $H_2/\mathopen{<} T_{6a}\mathclose{>}$ and  $D_3$ are isomorphic to each other.

In conclusion, the symmetry breaking pattern is $O_h\rightarrow D_3$.
Since $|O_h/D_3|=8$, the ground state degeneracy is eight, which is consistent with Eq. (\ref{eq:M12}).

\subsubsection{The $M_I$ phase: $O_h\rightarrow D_2$}

In Ref. \onlinecite{Luo2021b}, it is shown that there is an $M_I$ phase in the phase diagram of the spin-1 bond-alternating Kitaev-Gamma model, in which the spin alignments have six-site periodicity.
We will demonstrate that the unbroken symmetry group for the $M_I$ phase is isomorphic to the $D_2$ group in the sense of modulo $T_{6a}$.
Hence, the symmetry breaking pattern is $O_h\rightarrow D_2$, corresponding to $12$-fold ground state degeneracy since $|O_h/D_2|=12$. 
Using the residual unbroken $D_2$ symmetry group, the spin alignments in one of the twelve degenerate ground states can be determined as
\begin{flalign}
&\langle \vec{S}_{1+6n}\rangle = \left(\begin{array}{c}
c_1\\
0\\
c_2
\end{array}\right),
\langle \vec{S}_{2+6n}\rangle = \left(\begin{array}{c}
a\\
0\\
b
\end{array}\right),
\langle \vec{S}_{3+6n}\rangle = \left(\begin{array}{c}
a^\prime\\
0\\
b^\prime
\end{array}\right),\nn\\
&\langle \vec{S}_{4+6n}\rangle = \left(\begin{array}{c}
b^\prime\\
0\\
a^\prime
\end{array}\right),
\langle \vec{S}_{5+6n}\rangle = \left(\begin{array}{c}
b\\
0\\
a
\end{array}\right),
\langle \vec{S}_{6+6n}\rangle = \left(\begin{array}{c}
c_2\\
0\\
c_1
\end{array}\right).
\label{eq:M_I}
\end{flalign}
It is straightforward to see that the center of mass spin direction of the six spins in a unit cell in Eq. (\ref{eq:M_I}) points to the middle of the edge $AB^\prime$.
More generally, the center of mass spin directions in the twelve degenerate ground states are directed at the middle points of the twelve edges of the spin cube in Fig. \ref{fig:Oh}.
We note that in Ref. \onlinecite{Luo2021b}, 
the spin alignments are identified to be given by Eq. (\ref{eq:M_I}), with  the relations $c_1=c_2$, $a=a^\prime$, and $b=b^\prime$.
However, these additional constraints are generally not satisfied according to the symmetry analysis. 

Consider the following group as the unbroken symmetry group,
\bea
H_3=\mathopen{<}TR(\hat{y},\pi),rt,T_{6a}\mathclose{>},
\eea
in which $rt$ can be worked out as
\bea
rt=R_{AB} IT_a,
\eea
where $R_{AB}$ is a $\pi$-rotation in the spin space defined as
\bea
R_{AB}:(S_i^x,S_i^y,S_i^z)\rightarrow (S_i^z,-S_i^y,S_i^x).
\eea
Clearly, the invariance of the spin orderings under $TR(\hat{y},\pi)$ requires that $S_i^y=0$.
On the other hand, the action of $R_{AB}IT_a$ is given by
\bea
R_{AB}IT_a:(S_i^x,S_i^y,S_i^z)\rightarrow (S_{7-i}^z,-S_{7-i}^y,S_{7-i}^x).
\eea
Therefore, within a six-site unit cell, the invariance under $R_{AB}IT_a$ leads to
\begin{flalign}
&S_1^x=S_6^z=c_1,~S_1^z=S_6^x=c_2,\nn\\
&S_2^x=S_5^z=a,~S_2^z=S_5^x=b,\nn\\
&S_3^x=S_4^z=a^\prime,~S_3^z=S_4^x=b^\prime,
\end{flalign}
which is consistent with  Eq. (\ref{eq:M_I}).

Next we prove that $H_3/\mathopen{<} T_{6a}\mathclose{>}\cong D_2$.
Let $a_3=rt$, $b_3=TR(\hat{y},\pi)$.
It can be easily seen that 
\bea
(a_3)^2=(b_3)^2=(a_3b_3)^2=1.
\eea
According to Eq. (\ref{eq:Dn}), this demonstrates that $H_3/\mathopen{<}T_{6a}\mathclose{>}$ is a subgroup of $D_2$.
In addition, there are at least four distinct group elements in $H_3/<T_{6a}>$, which are given by
\bea
\{1,a_3,b_3,a_3b_3\}.
\label{eq:elements_H3}
\eea 
The operations in Eq. (\ref{eq:elements_H3}) are all distinct, 
since when restricted to the spin space, they correspond to the four symmetry operations which leave invariant the line connecting the middle points of $AB^\prime$ and $A^\prime B$ in Fig. \ref{fig:Oh}.
Therefore, the order of $H_3/\mathopen{<} T_{6a}\mathclose{>}$ is no less than the order of $D_2$.
Hence $H_3/\mathopen{<} T_{6a}\mathclose{>}$ must be isomorphic to $D_2$.

\begin{figure}
\begin{center}
\includegraphics[width=8cm]{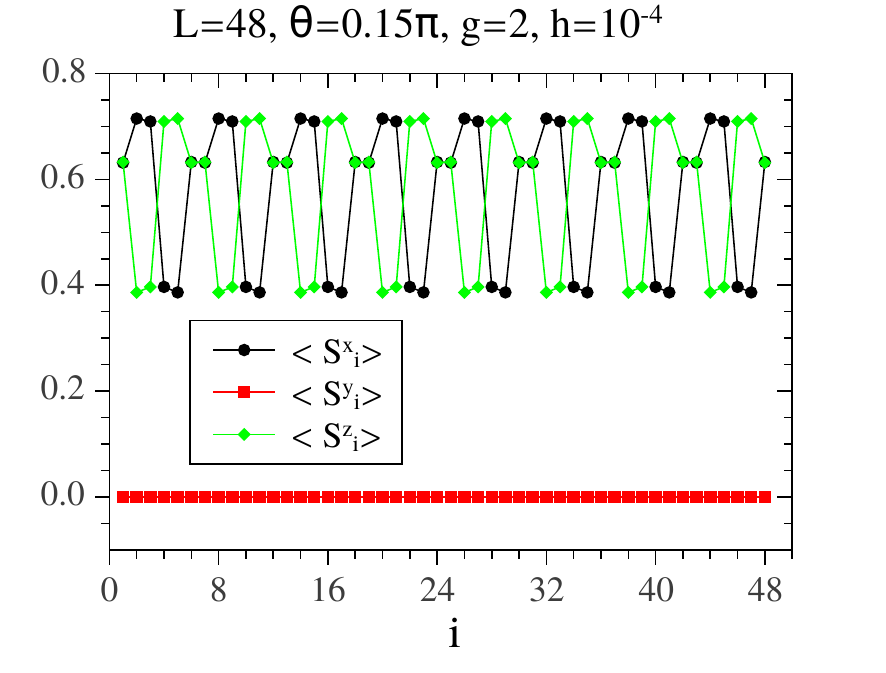}
\caption{Spin alignments in the spin-1 bond alternating Kitaev-Gamma chain at $\theta=0.15\pi$ and $g=2$, where $K=\sin(\theta)$, $\Gamma=\cos(\theta)$ and $g=g_y/g_x$. 
A small magnetic field $h=10^{-4}$ along $(1,0,1)$-direction is applied to polarize the system into the ground state in Eq. (\ref{eq:M_I}).
DMRG calculations are performed on a finite chain of $L=48$ sites with  periodic boundary conditions. 
} \label{fig:MI_phase}
\end{center}
\end{figure}

As shown in Fig. \ref{fig:MI_phase},  we have used DMRG numerics to calculate the spin expectation values for $\theta=0.15\pi$, $g=2$ in the spin-1 bond-alternating Kitaev-Gamma model,  which, according to Ref. \onlinecite{Luo2021b}, lies in the $M_I$ phase, where $K=\sin(\theta)$, $\Gamma=\cos(\theta)$ and $g=g_y/g_x$.
A small magnetic field $h=10^{-4}$ is applied along the $(1,0,1)$-direction, such that the state having spin alignments given in Eq. (\ref{eq:M_I}) is picked out among the twelve symmetry breaking ground states. 
We find that the  pattern of spin expectation values in Eq. (\ref{eq:M_I}) is satisfied, with 
\bea
c_1=0.6315&,& c_2=0.6326,\nn\\
a=0.7148&,& a^\prime=0.7093,\nn\\
b=0.3864&,&b^\prime=0.3968.
\eea
Notice that $c_1$, $c_2$ are very close, so are $a,a^\prime$ and $b,b^\prime$,
which is probably the reason why Ref. \onlinecite{Luo2021b} identify them to be equal.

\subsection{Dimerization order parameters}
\label{subsec:spontaneous}

Spontaneous dimerization can be induced in the spin chain when the system is coupled to phonons via spin-Peierls effect. 
It also arises in the $J_1$-$J_2$ models when the coupling of the next nearest neighbor interaction is beyond a critical value. 
In this subsection, we derive the order parameters for the spontaneous dimerization based on a symmetry analysis.

When the system spontaneously develops an alternation of strong and weak bonds, 
the symmetry breaking pattern is
$G_u \rightarrow  G$,
in which $G_u/\mathopen{<} T_{3a}\mathclose{>}\cong O_h$ and $G/\mathopen{<} T_{6a}\mathclose{>}\cong O_h$,
i.e., there is a $\mathbb{Z}_2$ symmetry breaking and the only broken symmetry is $T_{3a}$. 

The relations among dimerization order parameters can be determined using the unbroken symmetry group $G$.
First notice that the global $\mathbb{Z}_2\times \mathbb{Z}_2 (=\{1,R(\hat{x},\pi),R(\hat{y},\pi),R(\hat{z},\pi)\}$) symmetries are unbroken, hence all expectation values of cross products $\langle S_i^\alpha S_j^\beta \rangle$ ($\alpha\neq \beta$) vanish. 
Therefore, we investigate the following order parameters
\bea
O_{i}^{\alpha}= S_i^\alpha S_{i+1}^\alpha - S_{i+3}^\alpha S_{i+4}^\alpha.
\label{eq:dimer_order_def}
\eea
In one of the two dimerized states, since $T_{6a}$ is an unbroken symmetry, the expectation values of $O_{i}^{\alpha}$ satisfy
\bea
\langle O_{i}^{\alpha}\rangle =-\langle O_{i+3}^{\alpha}\rangle. 
\label{eq:O_relation_1}
\eea
Hence it is enough to consider $i\in\{6n+1,6n+2,6n+3\}$.
Furthermore, the actions of $R_a^{-1}T_{2a}$ and $R_IIT_a$ are given by
\bea
R_a^{-1}T_{2a}&:& \langle O_{i}^{\alpha} \rangle \rightarrow \langle O_{i+2}^{R_a^{-1}\alpha}\rangle ,\nn\\
R_IIT_a&:&\langle  O_{i}^{\alpha} \rangle \rightarrow   -\langle O_{3-i}^{R_I\alpha} \rangle,
\eea
which lead to the following relations
\begin{eqnarray}
A&=&\langle O_{1+6n}^{y}\rangle=\langle O_{1+6n}^{z}\rangle=-\langle O_{2+6n}^{x}\rangle\nn\\
&=&-\langle O_{2+6n}^{y} \rangle=\langle O_{3+6n}^{x} \rangle=\langle O_{3+6n}^{z}\rangle, \nn\\
B&=&\langle O_{1+6n}^{x} \rangle=-\langle O_{2+6n}^{z}\rangle=\langle O_{3+6n}^{y}\rangle.
\label{eq:relation_dimer_order_x}
\end{eqnarray}
An implication of Eqs. (\ref{eq:O_relation_1},\ref{eq:relation_dimer_order_x}) is 
\bea
\langle \vec{S}_i \cdot \vec{S}_{i+1} \rangle-\langle \vec{S}_{i+3} \cdot \vec{S}_{i+4} \rangle=(-)^{i-1}(4A+2B).
\eea
Notice that $\langle \vec{S}_i \cdot \vec{S}_{i+1} \rangle=\langle \vec{S}_{i+2} \cdot \vec{S}_{i+3} \rangle$ since $R_a^{-1}T_{2a}$ is a residual symmetry.
Therefore,  the conventional dimerization order parameter does not vanish:
\bea
(-)^i\langle \vec{S}_i \cdot \vec{S}_{i+1} \rangle+(-)^{i+1}\langle \vec{S}_{i+1} \cdot \vec{S}_{i+2} \rangle=-(4A+2B).
\eea

As an example, we consider the nearest neighbor spin-1/2 Kitaev-Gamma chain with an additional Heisenberg term in the six-sublattice rotated frame. 
The Hamiltonian is
\begin{eqnarray}
H_{nn}&=&\sum_{\langle ij \rangle\in\gamma\,\text{bond}} 
 \big[ -KS_i^\gamma S_j^\gamma -\Gamma (S_i^\alpha S_j^\alpha+S_i^\beta S_j^\beta)\big]\nn\\
 &&+\sum_{\langle\langle ij\rangle\rangle} J_2\vec{S}_i\cdot \vec{S}_j,
 \label{eq:H_nn}
\end{eqnarray}
in which $\langle\langle ij \rangle\rangle$ denotes a next nearest neighboring bond,
and the pattern for the bond $\gamma$ is shown in Fig. \ref{fig:bonds} (b).
We emphasize that $H_{nn}$ is not a realistic Hamiltonian describing real Kitaev materials, since the Heisenberg term acquires a weird form in the original frame. 
The system defined by Eq. (\ref{eq:H_nn}) is considered solely for the purpose of demonstrating spontaneous dimerization in a pure spin model without introducing the coupling to phonons.

\begin{figure*}[htbp]
\begin{center}
\includegraphics[width=8cm]{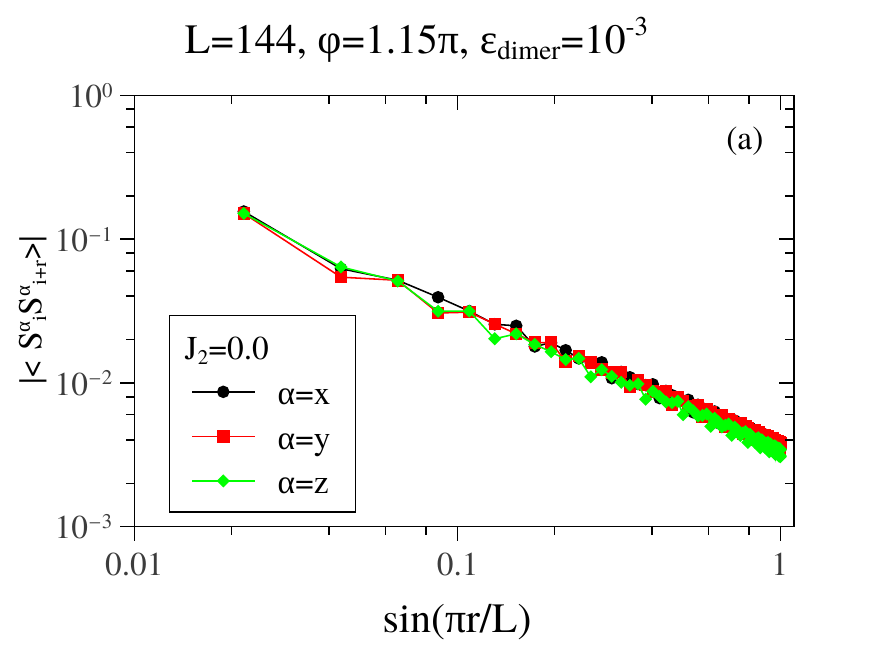}
\includegraphics[width=8cm]{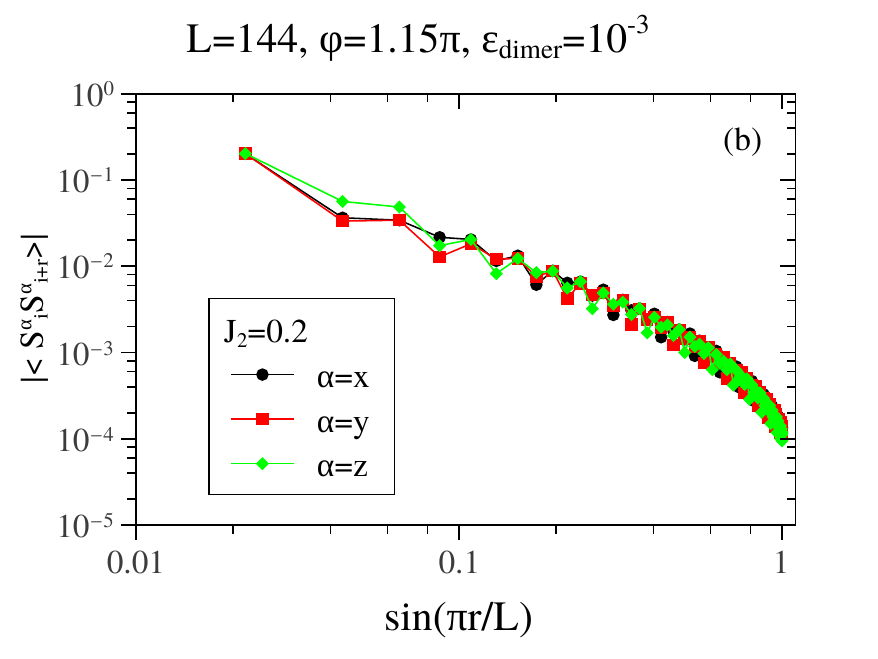}
\caption{Absolute values of the spin correlation functions $|\langle S_1^\alpha S_{1+r}^\alpha\rangle|$ ($\alpha=x,y,z$) at (a) $J_2=0$ and (b) $J_2=0.2$.
In both (a) and (b): the horizontal axes are $\sin(\pi r/L)$; the vertical and horizontal axes are plotted in logarithmic scales;
$\phi=0.15\pi$ where $K=\cos(\phi)$ and $\Gamma=\sin(\phi)$;
DMRG numerics are performed on a system of $L=144$ sites with periodic boundary conditions.
} \label{fig:dimer}
\end{center}
\end{figure*}

\begin{figure}
\begin{center}
\includegraphics[width=8cm]{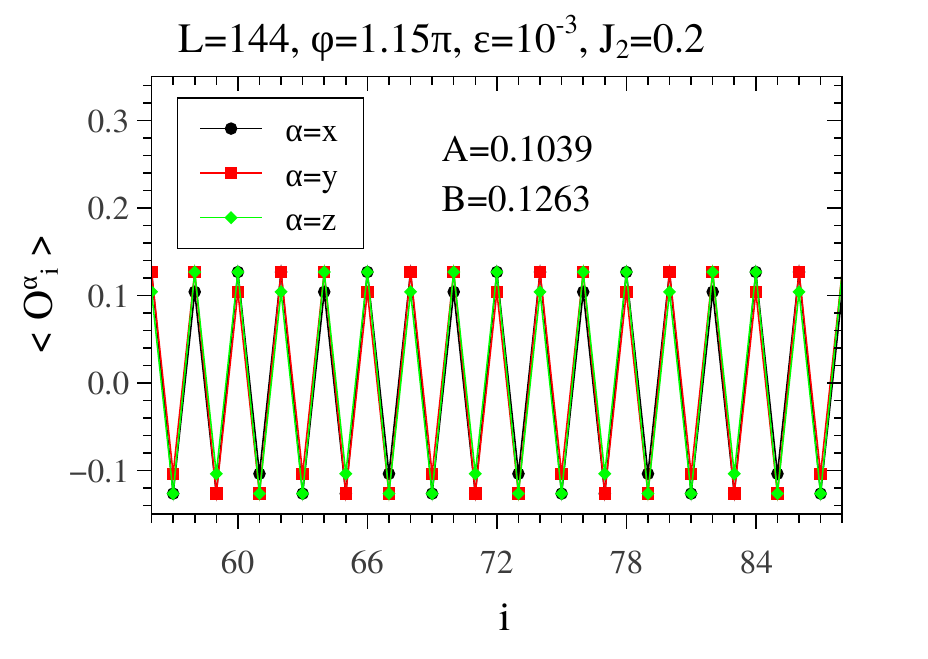}
\caption{Dimerization order parameters $\langle O_{i}^{\alpha}\rangle$ ($\alpha=x,y,z$) at $J_2=0.2$. 
$\phi=0.15\pi$ where $K=\cos(\phi)$ and $\Gamma=\sin(\phi)$, 
and DMRG numerics are performed on a system of $L=144$ sites with periodic boundary conditions.
Only a range of sites is shown for better illustrations of the patterns. 
} \label{fig:dimer2}
\end{center}
\end{figure}

As discussed in Ref. \onlinecite{Yang2020}, when $J_2=0$ in Eq. (\ref{eq:H_nn}),
there is an emergent SU(2)$_1$ phase in the phase diagram of the uniform spin-1/2 Kitaev-Gamma chain. 
The low energy theory is described by the SU(2)$_1$ Wess-Zumino-Witten (WZW) model, given by
\begin{eqnarray}
\mathcal{H}=\frac{2\pi}{3}v\int dx (\vec{J}_L\cdot \vec{J}_L+\vec{J}_R\cdot \vec{J}_R)-g_c\int dx \vec{J}_L\cdot \vec{J}_R,
\label{eq:Low_ham_KG}
\end{eqnarray}
in which $\vec{J}_\alpha$ ($\alpha=L,R$) are WZW current operators, $v$ is the velocity, and $g_c>0$ is the coupling of the marginal operator $\vec{J}_L\cdot \vec{J}_R$.
It is known that the marginal term is irrelevant (relevant) when $g_c>0$ ($<0$)\cite{Affleck1990}. 
In particular, the $\vec{J}_L\cdot \vec{J}_R$ term is irrelevant in the emergent SU(2)$_1$ phase of the spin-1/2 Kitaev-Gamma chain, though there are logarithmic corrections in the correlation functions due to its marginality \cite{Affleck1990,Yang2020}. 

When a nonzero $J_2$ is turned on in Eq. (\ref{eq:H_nn}),
the low energy theory remains  SU(2) symmetric.
Hence, the low energy Hamiltonian is still given by Eq. (\ref{eq:Low_ham_KG}),
albeit with renormalized values of $v$ and $g_c$.
When $g_c$ changes sign,  the  $\vec{J}_L\cdot \vec{J}_R$ term becomes marginally relevant.
As a result, $g_c$ flows to strong coupling in the low energy limit, giving rise to a dimerized phase \cite{Affleck1990}
with broken translational symmetry.
As an example, the $\phi=1.15\pi$ point lies in the emergent SU(2)$_1$ phase \cite{Yang2020},  
where $K=\cos(\phi)$ and  $\Gamma=\sin(\phi)$.
According to Supplementary Materials in Ref. \onlinecite{Yang2020},  the critical value of $J_{2}$ for $\phi=1.15\pi$ separating the gapless and dimerized phases
is $J_{c2}=0.135$. 
Hence, we expect that a spontaneous dimerization develops in the system when $J_2>J_{c2}$,
and the dimerization order parameters should satisfy the relations in Eq. (\ref{eq:relation_dimer_order_x}).

Next we proceed to numerical checks of the above symmetry and field theory predictions.
Fig. \ref{fig:dimer} (a) and (b) show the spin-spin correlation functions $\langle S_1^\alpha S_{1+r}^\alpha\rangle$ ($\alpha=x,y,z$) for $J_2=0$ and $J_2=0.2$, respectively, where $\phi=1.15\pi$. 
DMRG numerics are performed on a system of $L=144$ sites.
Clearly, the correlation functions exhibit a power law behavior when $J_2=0$, consistent with the predicted Luttinger liquid behavior. 
However, when $J_2=0.2$, the correlations decay exponentially at long distances, indicating an absence of rank-1 spin orders.

Fig. \ref{fig:dimer2} displays the dimerization order parameters defined in Eq. (\ref{eq:dimer_order_def}) in the presumably dimerized phase at $J_2=0.2$.
A very small dimerization $\epsilon=10^{-3}$ is introduced into the system to polarize the ground state into one of the two degenerate dimerized states, where $\epsilon=(g_x-g_y)/(g_x+g_y)$.
It can be straightforwardly observed that the relations  in Eq. (\ref{eq:relation_dimer_order_x}) are satisfied, and the values of the dimerization order parameters can be extracted as
\bea
A=0.1039, ~ B=0.1263.
\eea

\subsection{Magnetic field along $(1,1,1)$-direction}
\label{subsec:KG_field}

In this subsection, we discuss the symmetry group of the system when a magnetic field is applied along the $(1,1,1)$-direction. 
According to Eq. (\ref{eq:6rotation}), the uniform  magnetic field becomes staggered after six-sublattice rotation. 
The Hamiltonian in the six-sublattice rotated frame is given by
\begin{flalign}
&H^\prime_h=\sum_{<ij>\in\gamma\,\text{bond}} (g_0+(-)^{i-1}\delta) \big[ -KS_i^\gamma S_j^\gamma\nn\\
&-\Gamma (S_i^\alpha S_j^\alpha+S_i^\beta S_j^\beta)\big]+h\sum_i (-)^i (S_i^x+S_i^y+S_i^z).
\label{eq:H6_2}
\end{flalign}

The remaining symmetries of $H_h^\prime$ among the operations in Eq. (\ref{eq:symmetry}) are $R_a^{-1} T_{2a}$ and $R_I I T_a$.
Hence, the symmetry group $G_h$ is generated by these two operations, i.e.,
\begin{eqnarray}
G_h=\mathopen{<}R_a^{-1} T_{2a},R_I I T_a\mathclose{>}.
\end{eqnarray}
We will prove that 
\begin{eqnarray}
G_h/\mathopen{<} T_{6a}\mathclose{>}\cong D_3.
\end{eqnarray}

Denote $a_h=R_a^{-1}T_{2a}$, and $b_h=R_IIT_a$.
It is straightforward to verify that 
\begin{eqnarray}
(a_h)^3=T_{6a},~ (b_h)^2=1,~(a_hb_h)^2=1.
\end{eqnarray}
Since $G_h/\mathopen{<} T_{6a}\mathclose{>}$ satisfies the generator-relation representation of the $D_3$ group,
it must be a subgroup of $D_3$.
On the other hand, the following six elements in $G_h/\mathopen{<} T_{6a}\mathclose{>}$ are all distinct 
\begin{eqnarray}
\{1,a_h,(a_h)^2,b_h,b_ha_h,b_h(a_h)^2\}.
\label{eq:elements_Gh}
\end{eqnarray}
This is because the operations in Eq. (\ref{eq:elements_Gh}), when restricted to the spin space,  correspond to the six symmetry operations of the green dashed triangle in Fig. \ref{fig:Oh}.
Hence $|G_h/\mathopen{<} T_{6a}\mathclose{>}|\geq |D_3|$. 
Combining the established fact that $G_h/\mathopen{<} T_{6a}\mathclose{>}$ is a subgroup of $D_3$, we conclude that $G_h/\mathopen{<} T_{6a}\mathclose{>}$ is isomorphic to $D_3$.

We also note that the symmetry group $G_h$ is  nonsymmorphic, which cannot be written as a semi-direct product structure as $D_3\ltimes \mathopen{<}T_{6a}\mathclose{>}$.
This is because $R_a^{-1}T_{2a}$ remains to be an element in $G_h$, so that the analysis in  Eq. (\ref{eq:exact_C3}) still applies.

\section{Bond-alternating Kitaev-Heisenberg-Gamma chain}
\label{sec:KHG}

In this section, we generalize the previous symmetry analysis to bond-alternating Kitaev-Heisenberg-Gamma spin chain.
We also briefly discuss more general cases, including the Kitaev-Heisenberg-Gamma-$\Gamma^\prime$ model, 
and interactions beyond the nearest neighbor level.

\subsection{Kitaev-Heisenberg-Gamma chain}

The symmetry group $G_{1u}$ of a uniform Kitaev-Heisenberg-Gamma (KH$\Gamma$) spin chain has been shown in Ref. \onlinecite{Yang2021} to satisfy $G_{1u}/\mathopen{<} T_{3a}\mathclose{>}\cong D_{3d}$.
In this section, we discuss the bond-alternating KH$\Gamma$ spin chain, and demonstrate that its symmetry group $G_{1}$ satisfies  $G_{1}/\mathopen{<} T_{6a}\mathclose{>}\cong D_{3d}$.

The Hamiltonian is defined as Eq. (\ref{eq:Ham}), 
in which $H^{(\gamma)}_{i,j}$ is replaced by
\begin{eqnarray}
H^{(\gamma)}_{i,j}=KS_i^{\gamma}S_j^{\gamma}+\Gamma (S_i^{\alpha}S_j^{\beta}+S_i^{\beta}S_j^{\alpha})+J\vec{S}_i\cdot \vec{S}_j.
\end{eqnarray}
It can be verified that after six-sublattice rotation, the Hamiltonian becomes
\begin{flalign}
&H^\prime=\sum_{<ij>\in\gamma\,\text{bond}} (g_0+(-)^{i-1}\delta) \big[ -(K+J)S_i^\gamma S_j^\gamma\nn\\
&-\Gamma (S_i^\alpha S_j^\alpha+S_i^\beta S_j^\beta)
-J(S_i^\alpha S_j^\beta+S_i^\beta S_j^\alpha)\big].
\label{eq:Ham6_b}
\end{flalign}
Compared with Eq. (\ref{eq:symmetry}), the elements in the $\mathbb{Z}_2\times \mathbb{Z}_2$ subgroup $\{1,R(\hat{x},\pi),R(\hat{y},\pi),R(\hat{z},\pi)\}$ -- except the identity -- are no longer symmetries of the model. 
As discussed in Sec. \ref{subsec:KG_field}, the symmetry group generated by $\{R_a^{-1}T_{2a},R_IIT_a\}$ modulo $T_{6a}$ is $D_3$.
Hence, in the sense of modulo $T_{6a}$, the symmetry group $G_{1}$ of the bond-alternating KH$\Gamma$ spin chain is
\begin{eqnarray}
G_{1}/\mathopen{<}T_{6a}\mathclose{>}\cong D_3\times \mathbb{Z}_2^T(\cong D_{3d}),
\end{eqnarray}
since $T$ acts as inversion in the spin space. 
Again,  $R_a^{-1}T_{2a}$ is a group element of $G_1$, and $G_1$ is nonsymmorphic as a consequence of the  short exact sequence in Eq. (\ref{eq:exact_C3}).

We make a comment on the relations between the symmetry operations in the rotated and un-rotated frames. 
As can be easily seen from Fig. \ref{fig:bonds} (a), the symmetry group of the Hamiltonian in the un-rotated frame for the bond-alternating KH$\Gamma$ model is generated by $T$, $T_{2a}$ and $IT_a$. 
By straightforward calculations, it can be shown that
\begin{eqnarray}
U_6^{-1} R_aT_a U_6=T_{2a}, ~ U_6^{-1}R_IIT_aU_6=IT_a.
\label{eq:sym_original}
\end{eqnarray}
Hence the symmetry groups in the rotated and unrotated frames just differ by a $U_6$ transformation, which is as expected. 
However, we note that the $\mathbb{Z}_2\times \mathbb{Z}_2$ symmetry for the bond-alternating Kitaev-Gamma model acquires a complicated form with six-site periodicity in the original frame, as discussed in Ref. \onlinecite{Yang2020}.

\subsection{$\Gamma^\prime$ term and interactions beyond nearest neighbor}

The $\Gamma^\prime$ term on the bond $\gamma\in\{x,y\}$ in the original frame is defined as
\bea
\Gamma^\prime (S_i^\gamma S_j^\alpha+S_i^\alpha S_j^\gamma+S_i^\gamma S_j^\beta+S_i^\beta S_j^\gamma),
\eea
which should be included in Eq. (\ref{eq:H_gamma}).
In addition to time reversal symmetry, it is straightforward to see that   $T_{2a}$  and $IT_a$ in Eq. (\ref{eq:H_gamma}) remain to be the symmetries even when a nonzero $\Gamma^\prime$ term is added. 
Therefore, comparing with the Kitaev-Heisenberg-Gamma chain, the symmetry group of a bond-alternating Kitaev-Heisenberg-Gamma-$\Gamma^\prime$ chain in the six-sublattice rotated frame remains to be $D_{3d}$ in the sense of modulo $\mathopen{<} T_{6a}\mathclose{>}$.

Finally, we consider interactions beyond the nearest neighbor level. 
At $n$'th neighbor level ($n\geq 2$), the couplings correspond to a $z$-bond\cite{Wang2017},
i.e., $\gamma=z$ in Eq. (\ref{eq:H_gamma}) where $j=i+n$.
Again, the bond structure is invariant under both $T_{2a}$  and $IT_a$ in the original frame. 
Hence, the symmetry group remains to be isomorphic to $D_{3d} $ modulo $\mathopen{<} T_{6a}\mathclose{>}$.

\section{Kitaev spin ladders}
\label{sec:ladder}

\begin{figure}[h]
\begin{center}
\includegraphics[width=8.5cm]{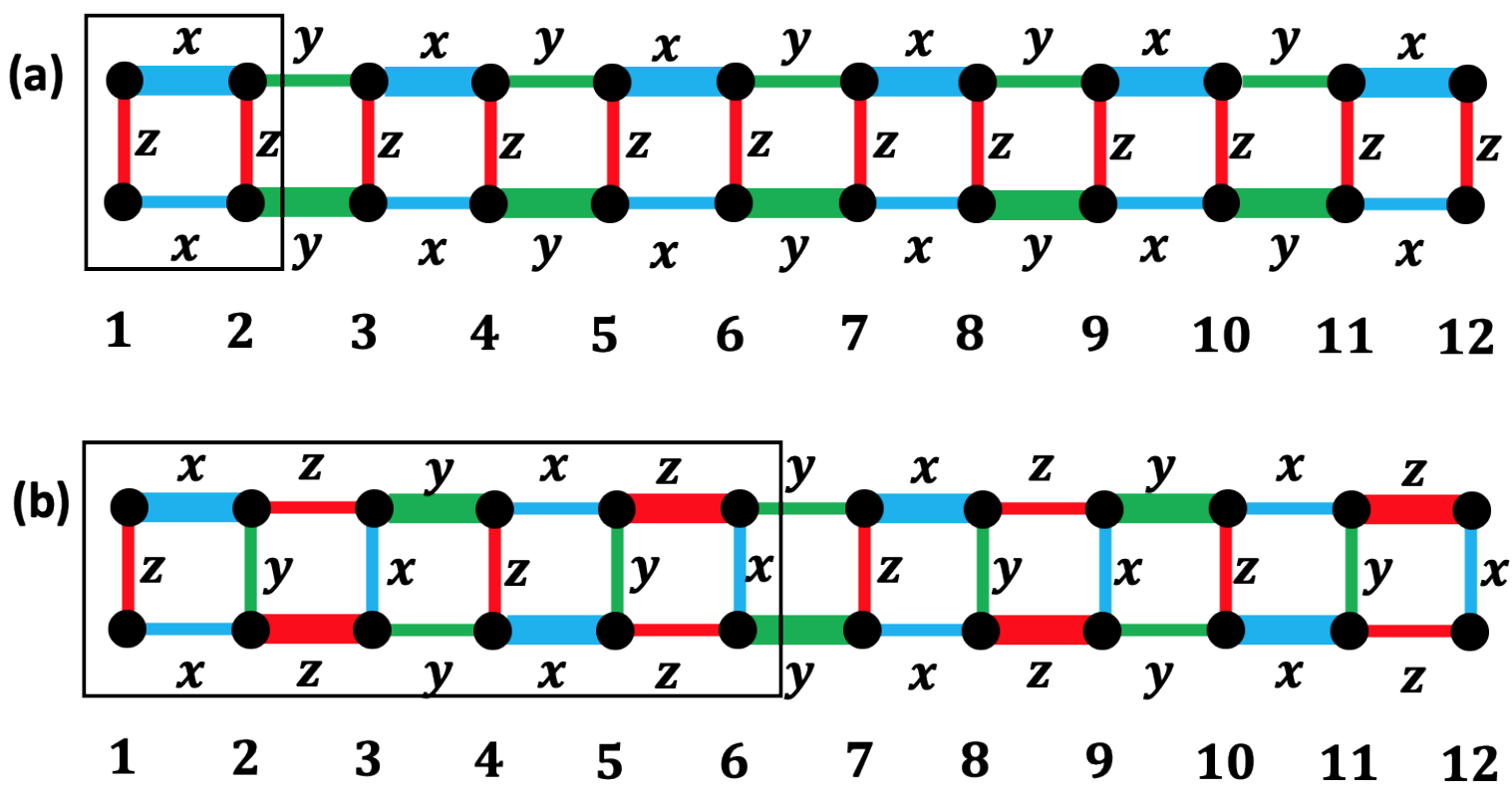}
\caption{The bond patterns of the bond-alternating generalized Kitaev spin ladder (a) before the six-sublattice rotation,
(b) after the six-sublattice rotation.
The thick colored lines represent bonds with anisotropy $g_x$, and the thin lines represent $g_y$.
} \label{fig:bonds_ladder}
\end{center}
\end{figure}

In this section, we  discuss the symmetry group structure of two-leg Kitaev spin ladders. 

\subsection{Bond-alternating Kitaev spin ladders}

The bond-alternating generalized Kitaev-Gamma two-leg spin ladder is shown in Fig. \ref{fig:bonds_ladder}.
In the ladder case, the six-sublattice rotation defined in Eq. (\ref{eq:6rotation})  is performed for both the upper and lower chains in Fig. \ref{fig:bonds_ladder} (a), and it can be verified that the bond pattern after the transformation is given by Fig. \ref{fig:bonds_ladder} (b).
It is clear that all the symmetries in the chain remain to be symmetry operations of the ladder, except that the inversion $IT_a$ for the chain should be replaced by reflection $M$ for the ladder, where $M$ is defined as
\begin{eqnarray}
M: \vec{S}_{+,i}\rightarrow \vec{S}_{+,7-i},~\vec{S}_{-,i}\rightarrow \vec{S}_{-,7-i},
\end{eqnarray}
in which  the subscripts $+$ and $-$ represent the upper and lower chains in the ladder, respectively. 
In addition to the symmetries inherited from  the chain case, there is an additional $\mathbb{Z}_2$ symmetry for the ladder, defined as
\begin{flalign}
&R_I P:\nn\\
& (S_{\lambda,i}^x,S_{\lambda,i}^y,S_{\lambda,i}^z)\rightarrow (-S_{-\lambda,10-i}^z,-S_{-\lambda,10-i}^y,-S_{-\lambda,10-i}^x),
\end{flalign}
in which $\lambda=\pm$.
This is a composition of an inversion with respect to the rung center at the fifth rung,  followed by a spin rotation $R_I$. 

The above analysis shows that the symmetry group $G_L$ is 
\begin{flalign}
G_L=\mathopen{<}T,R_a^{-1}T_{2a},R_IM,R(\hat{x},\pi),R(\hat{y},\pi),R(\hat{z},\pi),R_IP\mathclose{>}.
\end{flalign}
In the sense of modulo $T_{6a}$, the symmetry group is a semi-direct product of $O_h$ and $\mathopen{<}R_IP\mathclose{>}$, i.e.,
\begin{eqnarray}
G_L/\mathopen{<} T_{6a}\mathclose{>}=\mathbb{Z}_2\ltimes O_h.
\end{eqnarray}

The analysis can be  similarly performed for bond-alternating Kitaev-Heisenberg-Gamma two-leg spin ladder.
In this case, the structure  of the symmetry group $G_{1L}$ is 
\begin{eqnarray}
G_{1L}/\mathopen{<} T_{6a}\mathclose{>}\simeq \mathbb{Z}_2\ltimes D_{3d}.
 \label{eq:ladder_dimer_sym}
\end{eqnarray}

We note that both $G_L$ and $G_{1L}$ are nonsymmorphic which cannot be written as a semi-direct product involving $T_{6a}$, as a consequence of the short exact sequence in Eq. (\ref{eq:exact_C3}).

\subsection{Uniform Kitaev spin ladders}

Finally, we determine the symmetry group of Kitaev spin ladders without bond-alternation.
In the absence of dimerization, the swapping of the two chains is a symmetry, i.e.,
\begin{eqnarray}
\sigma:\vec{S}_{\lambda,i}\rightarrow \vec{S}_{-\lambda,i}.
\end{eqnarray}
Since $\sigma$ commutes with all other symmetry operations, we see that for the uniform Kitaev-Gamma spin ladder,
the structure of the  symmetry group $G_{uL}$  is 
\begin{eqnarray}
G_{uL}/\mathopen{<} T_{3a}\mathclose{>} \simeq \mathbb{Z}_2 \times O_h,
\end{eqnarray}
and for the uniform Kitaev-Heisenberg-Gamma spin ladder, the symmetry group $G_{1uL}$ is
\begin{eqnarray}
G_{1uL}/\mathopen{<} T_{3a}\mathclose{>}\simeq \mathbb{Z}_2 \times D_{3d},
\label{eq:ladder_nondimer_sym}
\end{eqnarray}
in which for both cases, $\mathbb{Z}_2$ represents $\mathopen{<}\sigma\mathclose{>}$.

Both $G_{uL}$ and $G_{1uL}$ are nonsymmorphic groups, which cannot be written as a semi-direct product involving $T_{3a}$.
This time, instead of $R_a^{-1}T_{2a}$, we consider the symmetry operation $R_aT_a$ and the following short exact sequence
\bea
1\xrightarrow[]{} \mathopen{<} T_{3a}\mathclose{>} \xrightarrow[]{i} \mathopen{<}R_aT_{a}\mathclose{>} \xrightarrow[]{\pi} C_3 \rightarrow 1.
\label{eq:exact_C3_b}
\eea
In Eq. (\ref{eq:exact_C3_b}), 
it can be proved in a similar way as Sec. \ref{sebsec:group_extension} that
there does not exist a group homomorphic $\tau$ such that $\pi\cdot\tau=id$,
which establishes the nonsymmorphic group structures of $G_{uL}$ and $G_{1uL}$.

\section{Conclusions}
\label{sec:conclusion}

In conclusion, we have analyzed the symmetry group structures for several one-dimensional  generalized Kitaev spin models with bond alternations. 
As applications of the symmetry analysis,  the symmetry breaking patterns of several magnetically ordered phases are determined in the bond-alternating Kitaev-Gamma spin chains, including $O_h\rightarrow D_4$, $O_h\rightarrow D_3$, and $O_h\rightarrow D_2$ symmetry breaking. 
The dimerization order parameters are also derived for the spontaneous dimerization. 
Our work is useful in understanding magnetic phases in related models   and may provide  guidance for the symmetry classifications of mean field solutions in further investigations.

{\it Acknowledgments}
W.Y. and I.A. acknowledge support from NSERC Discovery Grant 04033-2016.
W.Y., P.H. and R.R. acknowledge the support  from the Canada First Research Excellence Fund, Quantum Materials and Future Technologies Program.
A.N. acknowledges computational resources and services provided by Compute Canada and Advanced Research Computing at the University of
British Columbia. A.N. acknowledges support from the Max Planck-UBC-UTokyo Center for Quantum Materials and the Canada First Research Excellence Fund
(CFREF) Quantum Materials and Future Technologies Program of the Stewart Blusson Quantum Matter Institute (SBQMI).

\appendix

\section{Hamiltonian in rotated frame}
\label{app:6rotation}

The Hamiltonian for the bond-alternating Kitaev-Gamma model acquires the following form in the six-sublattice rotated frame,
\begin{widetext}
\begin{eqnarray}
H^\prime_{6n+1,6n+2}&=&g_x\big[-KS_{6n+1}^xS_{6n+2}^x
-\Gamma (S_{6n+1}^yS_{6n+2}^y+S_{6n+1}^zS_{6n+2}^z)\big],\nn\\
H^\prime_{6n+2,6n+3}&=&g_y\big[-KS_{6n+2}^zS_{6n+3}^z
-\Gamma (S_{6n+2}^xS_{6n+3}^x+S_{6n+2}^yS_{6n+3}^y)\big],\nn\\
H^\prime_{6n+3,6n+4}&=&g_x\big[-KS_{6n+3}^yS_{6n+4}^y
-\Gamma (S_{6n+3}^zS_{6n+4}^z+S_{6n+3}^xS_{6n+4}^x)\big],\nn\\
H^\prime_{6n+4,6n+5}&=&g_y\big[-KS_{6n+4}^xS_{6n+5}^x
-\Gamma (S_{6n+4}^yS_{6n+5}^y+S_{6n+4}^zS_{6n+5}^z)\big],\nn\\
H^\prime_{6n+5,6n+6}&=&g_x\big[-KS_{6n+5}^zS_{6n+6}^z
-\Gamma (S_{6n+5}^xS_{6n+6}^x+S_{6n+5}^yS_{6n+6}^y)\big],\nn\\
H^\prime_{6n+6,6n+7}&=&g_y\big[-KS_{6n+6}^yS_{6n+7}^y
-\Gamma (S_{6n+6}^zS_{6n+7}^z+S_{6n+6}^xS_{6n+7}^x)\big].
\end{eqnarray}
\end{widetext}

The Hamiltonian for the bond-alternating Kitaev-Heisenberg-Gamma model in the rotated frame is given by
\begin{widetext}
\begin{eqnarray}
H^\prime_{6n+1,6n+2}&=&g_x\big[-(K+J)S_{6n+1}^xS_{6n+2}^x-\Gamma (S_{6n+1}^yS_{6n+2}^y+S_{6n+1}^zS_{6n+2}^z)-J(S_{6n+1}^yS_{6n+2}^z+S_{6n+1}^zS_{6n+2}^y)\big],\nn\\
H^\prime_{6n+2,6n+3}&=&g_y\big[-(K+J)S_{6n+2}^zS_{6n+3}^z-\Gamma (S_{6n+2}^xS_{6n+3}^x+S_{6n+2}^yS_{6n+3}^y)-J(S_{6n+2}^xS_{6n+3}^y+S_{6n+2}^yS_{6n+3}^x)\big],\nn\\
H^\prime_{6n+3,6n+4}&=&g_x\big[-(K+J)S_{6n+3}^yS_{6n+4}^y-\Gamma (S_{6n+3}^zS_{6n+4}^z+S_{6n+3}^xS_{6n+4}^x)-J(S_{6n+3}^xS_{6n+4}^z+S_{6n+3}^zS_{6n+4}^x)\big],\nn\\
H^\prime_{6n+4,6n+5}&=&g_y\big[-(K+J)S_{6n+4}^xS_{6n+5}^x-\Gamma (S_{6n+4}^yS_{6n+5}^y+S_{6n+4}^zS_{6n+5}^z)-J(S_{6n+4}^yS_{6n+5}^z+S_{6n+4}^zS_{6n+5}^y)\big],\nn\\
H^\prime_{6n+5,6n+6}&=&g_x\big[-(K+J)S_{6n+5}^zS_{6n+6}^z-\Gamma (S_{6n+5}^xS_{6n+6}^x+S_{6n+5}^yS_{6n+6}^y)-J(S_{6n+5}^xS_{6n+6}^y+S_{6n+5}^yS_{6n+6}^x)\big],\nn\\
H^\prime_{6n+6,6n+7}&=&g_y\big[-(K+J)S_{6n+6}^yS_{6n+7}^y-\Gamma (S_{6n+6}^zS_{6n+7}^z+S_{6n+6}^xS_{6n+7}^x)-J(S_{6n+6}^xS_{6n+7}^z+S_{6n+6}^zS_{6n+7}^x)\big].
\end{eqnarray}
\end{widetext}

\section{Verification of Eq. (\ref{eq:Generator_Oh})}
\label{app:verify}

Before proceeding on, we fix some notations.
Let $\mathcal{R}$ be a rotation in spin space defined as 
\begin{eqnarray}
(\mathcal{R}(S^x),\mathcal{R}(S^y),\mathcal{R}(S^z))=(S^x,S^y,S^z)R,
\label{eq:spin_act}
\end{eqnarray}
in which  $R$ is a $3\times 3$ orthogonal matrix corresponding to $\mathcal{R}$.
Let $\mathcal{R}^\prime$ be another rotation with $R^\prime$ the corresponding matrix.
Then the composition $\mathcal{R}\mathcal{R}^\prime$ is given by
\begin{eqnarray}
\mathcal{R}\mathcal{R}^\prime: (S^x,S^y,S^z)\rightarrow (S^x,S^y,S^z) RR^\prime.
\label{eq:composition}
\end{eqnarray}
For later convenience, recall that $R_a=R(\hat{n}_a,-2\pi/3)$ and $R_I=R(\hat{n}_I,\pi)$ satisfy
\begin{eqnarray}
R_a:&(S_i^x,S_i^y,S_i^z)&\rightarrow (S_i^z,S_i^x,S_i^y),\nn\\
R_I:&(S_i^x,S_i^y,S_i^z)&\rightarrow (-S_i^z,-S_i^y,-S_i^x),
\label{eq:RaRI}
\end{eqnarray}
in which $\hat{n}_a=\frac{1}{\sqrt{3}}(1,1,1)^T$ is parallel to the line of $OA$ in Fig. \ref{fig:Oh}, 
and $\hat{n}_I=\frac{1}{\sqrt{2}}(1,0,-1)^T$ is parallel to the line passing through the point that bisects the edge $CD^{\prime}$ and the point that bisects $C^{\prime}D$ in Fig. \ref{fig:Oh}.
Thus the matrix representations are
\begin{eqnarray}
R_a=\left(\begin{array}{ccc}
0&1&0\\
0&0&1\\
1&0&0
\end{array}
\right),~
R_I=\left(\begin{array}{ccc}
0&0&-1\\
0&1&0\\
-1&0&0
\end{array}
\right).
\end{eqnarray}
In addition to the spin rotations, the spatial operations act as
\begin{eqnarray}
T_a:& i\rightarrow i+1,\nn\\
I:&i\rightarrow 8-i. 
\label{eq:spatial_act}
\end{eqnarray}
Eqs. (\ref{eq:spin_act},\ref{eq:spatial_act}) enable us to calculate the actions of any symmetry operation.

We  first verify the relations $r^2=s^2=t^2=e$.
Firstly,
\begin{eqnarray}
r^2=T^2\cdot (IT_{-5a})^2\cdot (R_I)^2=1, 
\end{eqnarray}
since $T^2=1,IT_{na}I=T_{-na}$ and $(R_I)^2=[R(\hat{n}_I,\pi)]^2=R(\hat{n}_I,2\pi)=1$.
Secondly,
\begin{eqnarray}
s^2 & =& T^2\cdot (T_{2a} I T_{-a})^2 \cdot (R_a^{-1} R_I R_a)^2\nn\\
&=& T_{2a}IT_{a}  I T_{-a} \cdot [R(R_a^{-1}\hat{n}_I,\pi)]^2\nn\\
&=& T_{2a} T_{-a} T_{-a}\cdot R(R_a^{-1}\hat{n}_I,2\pi)\nn\\
&=&1,
\end{eqnarray}
in which $R_0 R(\hat{n},\theta) R_0^{-1}=R(R_0\hat{n},\theta)$ is used.
Finally for $t$, we obtain
\begin{eqnarray}
t^2=T^2 \cdot [R(\hat{y},\pi)]^2=1.
\end{eqnarray}
We note that we view all the operations as acting in the three-dimensional vector space spanned by $\{S^x,S^y,S^z\}$.
Hence although SO(3) acts projectively for the spin-1/2 case, we still have $R(\hat{n},2\pi)=1$ when acting on $\{S^x,S^y,S^z\}$.
Also $T^2=1$ in $\text{span}\{S^x,S^y,S^z\}$, though $T^2=-1$ for spin-1/2.

Using the expressions of $r,s,t$, we can work out the expressions of $rs,st,rt$, as
\begin{eqnarray}
rs &=& R_a^{-1}T_{2a},\nn\\
st & =& R(\hat{z},\frac{\pi}{2}) I T_{-3a}, \nn\\
rt & =&R([AB],\pi) IT_{-5a}.
\label{eq:relation_ops}
\end{eqnarray}
Their actions in the spin space are given by
\begin{eqnarray}
\begin{array}{  c  c  c  }
rs & (x,y,z)\rightarrow (y,z,x) & R(OA,\frac{2\pi}{3}) \\
st & (x,y,z)\rightarrow (y,-x,z) & R(\hat{z},\frac{\pi}{2})\\
rt &(x,y,z)\rightarrow (z,-y,x) & R([AB],\pi).
\end{array}
\end{eqnarray}
in which 
the second and the third columns give the actions in the spin space (where $S^\alpha$ is denoted as $\alpha$ for short)
and the geometrical meanings as symmetries of a cube in Fig. \ref{fig:Oh}, respectively,
and $[AB]$ represents the line passing through the point that bisects the edge $AB^{\prime}$ and the point that bisects $A^{\prime}B$ in Fig. \ref{fig:Oh}.
In obtaining Eq. (\ref{eq:relation_ops}), we have used the following identities,
\begin{eqnarray}
R_IR_a^{-1}R_IR_a&=&R_a^{-1}\nn\\
R_a^{-1}R_IR_aR(\hat{y},\pi)&=&R(\hat{z},\pi/2)\nn\\
R_IR(\hat{y},\pi)&=&R([AB],\pi).
\end{eqnarray}

Next we verify the relations $(rs)^3=(st)^4=(rt)^2=e$.
Firstly,
\begin{eqnarray}
(rs)^3&=&(R_a)^{-3}\cdot (T_{2a})^{3}\nn\\
&=& [R(\hat{n}_a,-2\pi/3) ]^{-3} \cdot T_{6a}\nn\\
&=& R(\hat{n}_a,-2\pi) \cdot T_{6a}\nn\\
&=& T_{6a},
\end{eqnarray}
in which $R_a=R(\hat{n}_a,-2\pi/3)$ is used,
and clearly $(rs)^3=e$ modulo $T_{6a}$.
Secondly,
\begin{eqnarray}
(st)^4&=& (IT_{-3a} )^4 \cdot [R(\hat{z},\pi/2)]^4\nn\\
&=& (IT_{-3a}I)T_{-3a}(IT_{-3a}I)T_{-3a} \cdot R(\hat{z},2\pi)\nn\\
&=& T_{3a}T_{-3a}T_{3a}T_{-3a}\cdot R(\hat{z},2\pi)\nn\\
&=&1.
\end{eqnarray}
Finally,
\begin{eqnarray}
(rt)^2&=&(IT_{-5a})^2\cdot [R([AB],\pi)]^2\nn\\
&=&1.
\end{eqnarray}
This proves that all the relations in Eq. (\ref{eq:Generator_Oh}) are satisfied.
Hence $<r,s,t>/\mathopen{<}T_{6a}\mathclose{>}$ is isomorphic to a subgroup of $O_h$.

\section{Verification of Eq. (\ref{eq:symmetry_from_rst})}
\label{app:verify_B}

(1) To verify the expression of $T$, using Eq. (\ref{eq:generators},\ref{eq:relation_ops}), we calculate
\begin{eqnarray}
&&(rs)^{-1}\cdot(st)^2\cdot r\cdot (st)\nn\\
&=& (T_{2a})^{-1}  (IT_{-3a})^2 IT_{-5a}  (IT_{-3a})  \nn\\
&&\cdot [R_a] [R(\hat{z},\pi/2)]^2 [R_I]  [R(\hat{z},\pi/2)]\cdot T\nn\\
&=&T.
\label{eq:timereverse}
\end{eqnarray}
The spatial part of Eq. (\ref{eq:timereverse}) can be verified to be 1.
Using Eq. (\ref{eq:RaRI}), $R(\hat{z},\pi):(x,y,z)\rightarrow (-x,-y,z)$, $R(\hat{z},\pi/2):(x,y,z)\rightarrow (y,-x,z)$,
and the composition rule Eq. (\ref{eq:composition}),
it is a straightforward calculation to verify that $R_aR(\hat{z},\pi)R_IR(\hat{z},\pi/2): (x,y,z)\rightarrow (x,y,z)$.
Thus $sr(st)^2r(st)$ is equal to $T$.

(2) To verify the expression of $R_IIT_{a}$, we calculate
\begin{eqnarray}
&&(st)^2\cdot r \cdot st\cdot  s\cdot (rs)^3\nn\\
&=& (IT_{-3a})^2 (IT_{-5a}) (IT_{-3a}) (T_{2a}IT_{-a}) (T_{2a})^3 \nn\\
&&\cdot [R(\hat{z},\pi/2)]^2 [R_I][R(\hat{z},\pi/2)][R_a^{-1}R_IR_a][(R_a^{-1})^3]\cdot T^2\nn\\
&=&R_I IT_a.
\end{eqnarray}
The spatial part is $IT_a$, and $T^2=1$.
For the spin part, we can use Eq. (\ref{eq:composition}) and perform a matrix multiplication.
The result is
\begin{flalign}
&[R(\hat{z},\pi/2)]^2 R_IR(\hat{z},\pi/2)(R_a^{-1}R_IR_a)=\left(\begin{array}{ccc}
0&0&-1\\
0&-1&0\\
-1&0&0
\end{array}\right),
\end{flalign}
which is exactly $R_I$.

(3) To verify $R_a^{-1}T_{2a}$, we calculate
\begin{eqnarray}
r\cdot s&=&R_a^{-1}T_{2a},
\end{eqnarray}
which has been demonstrated in Eq. (\ref{eq:relation_ops}).

(4) To verify $R(\hat{x},\pi)$, we calculate
\begin{eqnarray}
&&rt\cdot st\cdot s\cdot r\nn\\
&=&(IT_{-5a})(IT_{-3a})(T_{2a}IT_{-a})(IT_{-5a})\nn\\
&&\cdot [R([AB],\pi)][R(\hat{z},\pi/2)][R_a^{-1}R_IR_a][R_I]\cdot T^2\nn\\
&=&R(\hat{x},\pi).
\end{eqnarray}
The spatial part can be verified to be 1.
By matrix multiplication, the spin part can be calculated to be  exactly $R(\hat{x},\pi)$.

(4) To verify $R(\hat{y},\pi)$, we calculate
\begin{eqnarray}
&&s\cdot r\cdot (st)^2\cdot rs\nn\\
&=& (T_{2a}IT_{-a})(IT_{-5a})(IT_{-3a})^2(T_{2a})\nn\\
&&\cdot [R_a^{-1}R_IR_a] [R_I] [R(\hat{z},\pi/2)]^2[R_a^{-1}]\cdot T^2\nn\\
&=& R(\hat{y},\pi).
\end{eqnarray}
Again, it can be verified that the spatial component is 1 and  the spin component is $R(\hat{y},\pi)$.

(5) To verify $R(\hat{z},\pi)$, we calculate
\begin{eqnarray}
&&(st)^2\nn\\
&=&(IT_{-3a})^2\cdot [R(\hat{z},\pi/2)]^2\nn\\
&=& R(\hat{z},\pi).
\end{eqnarray}

\section{Group extensions and second cohomology group}
\label{app:cohomology}

In this appendix, we briefly review the relation between the group extensions and the second cohomology group \cite{Dummit1991}.

We start from the short exact sequence in Eq. (\ref{eq:short_exact}).
Let $f$ be a function from $H\times H$ to $N$.
$f$ is called a $2$-cocycle if 
\bea
f(g,h)f(gh,k)=\varphi_g(f(h,k))f(g,hk),
\eea
and a $2$-coboundary if there exists a function $c: H\rightarrow N$ such that 
\bea
f(g,h)=c(g)\varphi_g(c(h))[c(gh)]^{-1}.
\label{eq:2coboundary}
\eea
It is straightforward to verify that a $2$-coboundary is also a $2$-cocycle.
By point-wise multiplications, the collections of $2$-cocycles and $2$-coboundaries form abelian groups $Z^2(H,N)$ and $B^2(H,N)$.
The second cohomology group of $H$ with coefficients in $N$ is defined by the quotient group
\bea
H^2(H,N)=Z^2(H,N)/B^2(H,N).
\label{eq:H2}
\eea

Consider an injective map $x:H\rightarrow G_0$, such that $\pi\cdot x=id$.
For any $g,h\in H$, since $\pi(x(g)x(h)x(gh)^{-1})=1$, we see that $x(g)x(h)x(gh)^{-1}$ has to be an element in $N=\text{Ker}(\pi)$.
Therefore, a function $f:H\times H \rightarrow N$  can be defined from the map $x$ as $f(g,h)=x(g)x(h)[x(gh)]^{-1}$.
It can be verified that $f$ defined in this way satisfies the $2$-cocycle condition due to the associativity of group multiplication.
Furthermore, we do not want to distinguish among different $f$-functions if the values of their corresponding $x$-maps only differ by elements in $N$.
That is to say,  the two $f$-functions defined from $xc$ and $x$ should be viewed as equivalent for any $c:H\rightarrow N$.
It can be verified that the difference between such two $f$-functions is exactly given by the expression in Eq. (\ref{eq:2coboundary}).
Therefore, the equivalent classes of the $f$-functions built from the $x$-maps correspond to elements in the second cohomology group $H^2(H,N)$ in Eq. (\ref{eq:H2}).
This builds the relation between the exact sequence in Eq. (\ref{eq:short_exact})
and the second cohomology group $H^2(H,N)$.
Notice that $f\equiv 1$ if $x$ is a group homomorphism.
Hence, in some sense, $H^2(H,N)$ measures the extent to which the equivalent class of $x$ breaks the homomorphism property.



\end{document}